\documentclass[useAMS,usegraphicx,usenatbib]{mn2e}
\usepackage{times}

\graphicspath{{plots/}}

\title[MGC: Dust]{The Millennium Galaxy Catalogue: the $B$-band
attenuation of bulge and disc light and the implied cosmic dust and
stellar mass densities}

\author[Simon P.~Driver et al.]{Simon P.~Driver,$^{1}$\thanks{E-mail:
spd3@st-and.ac.uk}
Cristina~C.~Popescu,$^{2}$
Richard~J.~Tuffs,$^{2}$
Jochen~Liske,$^{3}$ \newauthor 
Alister~W.~Graham,$^{4}$
Paul~D.~Allen$^{1}$ and
Roberto~De~Propris$^{5}$ \\
$^1$SUPA\thanks{Scottish Universities Physics Alliance}, School of
Physics and Astronomy, University of St Andrews, North Haugh, St
Andrews, Fife, KY16 9SS\\
$^2$Max-Planck-Institut f\"ur Kernphysik, Saupfercheckweg 1, 69117 
Heidelberg, Germany\\
$^3$European Southern Observatory, Karl-Schwarzschild-Str.~2, 85748
Garching, Germany \\ 
$^4$Centre for Astrophysics and Supercomputing, Swinburne University
of Technology, Hawthorn, Victoria 3122, Australia \\
$^5$Cerro Tololo Inter-American Observatory, Casilla 603, La Serena, Chile}

\newcommand{\bmgc}{B_{\mbox{\tiny \sc MGC}}}

\begin{document}

\date{Accepted
...... Received .....}

\pagerange{\pageref{firstpage}--\pageref{lastpage}} \pubyear{2007}

\maketitle

\label{firstpage}

\begin{abstract}
Based on our sample of 10095 galaxies with bulge-disc decompositions
we derive the {\em empirical} $\bmgc$-band internal
attenuation--inclination relation for galaxy discs and their
associated central bulges. Our results agree well with the
independently derived dust models of Tuffs et al., leading to a direct
constraint on the mean opacity of spiral discs of $\tau_B^{\rm f} =
3.8 \pm 0.7$ (central face-on $\bmgc$-band opacity). Depending on
inclination, the $\bmgc$-band attenuation correction varies from $0.2$
-- $1.1$~mag for discs and from $0.8$ -- $2.6$~mag for bulges. We find
that, overall, $37$ per cent of all $\bmgc$-band photons produced in
discs in the nearby universe are absorbed by dust, a figure that rises
to $71$ per cent for bulge photons. {\em The severity of internal dust
extinction is such that one must incorporate internal dust
corrections in all optical studies of large galaxy samples. This is
particularly pertinent for optical {\it HST} comparative evolutionary
studies as the dust properties will also be evolving.} We use the new
results to revise our recent estimates of the spheroid and disc
luminosity functions. The implied stellar mass densities at redshift
zero are somewhat higher than our earlier estimates: $\rho_{\rm
discs} = (3.8 \pm 0.6) \rightarrow (4.4 \pm 0.6) \times 10^8 \,
h$~M$_{\odot}$~Mpc$^{-3}$ and $\rho_{\rm bulges} = (1.6 \pm 0.4)
\rightarrow (2.2 \pm 0.4) \times 10^8 \,
h$~M$_{\odot}$~Mpc$^{-3}$. From our best fitting dust models we derive
a redshift zero cosmic dust density of $\rho_{\rm dust} \approx (5.3
\pm 1.7) \times 10^5 \, h$~M$_{\odot}$~Mpc$^{-3}$. This implies that
$(0.0083 \pm 0.0027) \, h$ per cent of the baryons in the Universe are
in the form of dust and $(11.9 \pm 1.7) \, h$ per cent
(Salpeter-`lite' IMF) are in the form of stars ($\sim58$ per cent
reside in galaxy discs, $\sim10$ per cent in red elliptical galaxies,
$\sim29$ per cent in classical galaxy bulges and the remainder in low
luminosity blue spheroid systems/components).
\end{abstract}



\section{Introduction}
Internal dust attenuation of the photon flux from galaxies is a severe
issue at optical wavelengths, reducing the total emergent luminosity,
introducing an inclination dependence of the observed flux (e.g.,
Giovanelli et al.\ 1995; Masters et al.\ 2003)  and modifying the galaxy
light-profile shape. Model predictions of these effects have been
given by many authors, including Kylafis \& Bahcall 1987; Byun,
Freeman \& Kylafis 1994; Evans 1994; Kuchinski et al.\ 1998; Bianchi
et al.\ 1996; Ferrara et al.\ 1999; Baes \& Dejonghe 2001; Cunow 2001;
Tuffs et al.\ 2004; Pierini et al.\ 2004;
\citealp{mollenhoff06}). Typically, spiral galaxies exhibit a radial
gradient in opacity, with higher opacities in the central regions
(Boissier et al.\ 2004; Popescu et al.\ 2005. The dust is also
typically distributed in a thinner layer than the stellar
population(s) (Xilouris et al.\ 1999). Because of these two properties
of the dust distribution, the {\em observed} stellar profiles appear
less centrally concentrated (i.e., lower S\'ersic indices),
scale-lengths are overestimated, luminosities are underestimated and
the central surface brightnesses artificially dimmed. All of these
effects become more pronounced at higher inclinations\footnote{Here we
define inclination to run from face-on (low inclination, $i=0\deg$) to
edge-on (high inclination, $i=90\deg$).} (see \citealp{mollenhoff06}
for detailed model predictions). Dust may also exacerbate the issue of
inner disc truncation leading to Type II profiles (Freeman 1970) by
preferentially suppressing core flux.

Bulges, even though they are assumed to contain minimal dust, are
still seen through the dust layer of the disc, and therefore also
suffer attenuation of their stellar light and modification of their
surface-brightness distributions due to dust. In fact, these effects
are predicted to be even more pronounced for bulges than for the discs
(Tuffs et al.\ 2004), since bulges are concentrated towards the central
regions of galaxies, where the opacity is highest.

In terms of obtaining statistical distributions for the disc and bulge
components (e.g., disc/bulge luminosity--size or luminosity--colour
distributions, see \citealp{mgc06}), samples uncorrected for galaxy
inclination will have their intrinsic distributions broadened (and
skewed, see in particular the recent papers by Choi, Park \& Vogeley
2007; and Shao et al.\ 2007).
Indeed, recent observational studies have produced evidence that the
size distribution of galaxies, at fixed luminosity, is broader than
that predicted from hierarchical simulations (see Shen et al.\ 2003;
Driver et al.\ 2005). In terms of the total stellar mass the effect is
less obvious. This quantity is usually estimated by multiplying a
galaxy's luminosity by its stellar mass-to-light ratio, which is in
turn derived from the correlation between mass-to-light ratio and
colour (Bell \& de Jong 2001). The point is that the decrease in the
estimated stellar mass due to attenuation will be counteracted, at
least to some extent, by the increase in the stellar mass due to
reddening.  Bell \& de Jong (2001) argue that these two effects mostly
cancel (see their fig.~1). We find that while this is true for
moderately inclined systems ($i < 60\deg$) it does not hold for highly
inclined systems.

Historically, there has been some debate as to whether studies of
galaxies at various inclinations can be used to constrain the dust
distribution and opacity simultaneously (\citealp{holmberg58};
\citealp{ddp90}; \citealp{valentijn}; \citealp{disney92}). The general
consensus was that one could not (see \citealp{davies}) and
alternative paths to the dust distributions have now been pursued
(e.g., overlapping galaxies, Keel \& White 2001; surface
brightness--scale-length relations, Graham 2001; quasar sight-lines
through foreground galaxies, \"Ostman, Goobar \& Mortsell 2006; and
self-consistent modelling of the UV/optical/FIR/sub-mm emission from
galaxies, Silva et al.\ 1998; Bianchi et al.\ 2000; Popescu et al.\
2000; Popescu \& Tuffs 2004). However, with the dust models (including
3D distribution, clumpiness and grain composition) now constrained by
independent methods, it should be possible to revisit the use of large
statistical catalogues to constrain the mean opacity of discs.

In particular, given a complete galaxy sample with comprehensive
structural analysis and high-completeness redshift coverage one should
be able to determine, {\em empirically}, the attenuation--inclination
relation.\footnote{We note that attenuation is an integral property of
an extended distribution of light and should not be confused with the
extinction along a single line of sight.} This can be achieved if a
characteristic feature can be identified in the galaxy population and
this feature measured for sub-samples of varying inclination. One
obvious feature is the turn-over of the luminosity function, i.e.\
$L^*$ or $M^*$, which in the absence of dust should be inclination
independent. Using large galaxy samples drawn from contemporary
surveys $M^*$ can now be measured to an accuracy of $\Delta M^* <
0.1$~mag (see for example \citealp{zucca}; \citealp{norberg};
\citealp{sdsslf}; \citealp{mgc05}).

Predictions based on detailed dust models suggest that the total
$B$-band magnitude of a galaxy disc may be attenuated by up to $2$~mag
and that of a galaxy bulge by up to $2.5$~mag, depending on opacity
and viewing angle (see
Tuffs et al.\ 2004; Pierini et al.\ 2004). In this paper we report our
empirical estimate of the relationship between attenuation and
inclination, and use it to constrain the mean opacity of discs by
comparison to the model predictions of Tuffs et al.\ (2004; which in
turn are based on the dust model of Popescu et al.\ 2000).  Using a
combination of (i) our empirical correction to remove the inclination
dependent attenuation, and (ii) the dust model to evaluate and remove
the face-on attenuation, we revise the luminosity functions and total
stellar mass densities recently reported in Driver et al.\ (2007). We
also combine the mean opacity of discs and total disc stellar mass to
derive the cosmic dust density.

In Section~\ref{data} we review the data used in this investigation
and in Section~\ref{analysis} we describe our empirical analysis
leading to a constraint on the mean opacity of galaxies. In
Section~\ref{model} we use our results to constrain our adopted dust
model (Tuffs et al.\ 2004). In Section~\ref{cosmic} we recover the
spheroid and disc luminosity functions through a variety of methods
and discuss the implications of our findings in the context of the
cosmic baryon budget. Throughout we adopt the following cosmological
parameters: $\Omega_{\rm M}=0.3$, $\Omega_{\Lambda}=0.7$ and $H_0=100
\, h$~km~s$^{-1}$~Mpc$^{-1}$.

\section{The Millennium Galaxy Catalogue}
\label{data}
The Millennium Galaxy Catalogue (MGC) spans a $37.5$~deg$^2$ region of
the vernal equatorial sky and contains 10095 galaxies brighter than
$\bmgc = 20$~mag with $96$ per cent spectroscopic redshift
completeness. The imaging catalogue is described in Liske et al.\
(2003), the spectroscopic follow-up in Driver et al.\ (2005) and the
photometric accuracy and completeness in Cross et al.\ (2004), Driver
et al.\ (2005) and Liske et al.\ (2006). We have decomposed this
sample into bulges and discs with {\sc gim2d} (Simard et al.\ 2002),
using an $R^{1/n}$ S\'ersic profile for bulges and an exponential
profile for discs (Allen et al.\ 2006) and providing an extensive
bulge-disc resource which is publicly available at
http://www.eso.org/$\sim$jliske/mgc/. In this paper we use the MGC
structural catalogue {\sc mgc\_gim2d}. 

We point out that not all galaxies in our catalogue are two-component
systems. Allen et al.\ (2006) divided those objects that were best fit
with a single-component S\'ersic profile into pure discs and pure
bulges (i.e.\ ellipticals) according to their S\'ersic index. Note
that we use the term `spheroid' to mean both ellipticals and bulges,
and we reserve the term `bulge' exclusively for the central 3D
structure of a two-component system. We further separate our
ellipticals into classical (red) and new (blue), as well as our bulges
into classical and pseudo at $(u-r)_{\rm core}=2$~mag according to the
spheroidal colour bimodality found by Driver et al.\ (2007, see their
fig.~2). In Sections~\ref{analysis} and \ref{model} we will use the
classical bulges, but not the ellipticals which are assumed dust free,
as well as all of the discs, irrespective of whether they `contain' a
bulge or not.

The robustness of our catalogue has been quantified using independent
repeat observations and {\sc gim2d} decompositions of $682$ galaxies.
These duplicate observations originate from the overlap regions of
neighbouring MGC fields and hence they include the effects of varying
observing conditions. This comparison sample demonstrates that for
components with $M_B < -17$~mag our structural catalogue is accurate
to better than $\pm 0.1$~mag for bulges and $\pm 0.15$~mag for discs,
and that the disc inclination, $i$, has $\Delta \cos(i) \sim 0.05$
(see fig.~15 of Allen et al.\ 2006). In our analysis we use bin sizes
of $0.1$ for $\cos(i)$ and $0.5$~mag for luminosity. Hence the errors
are smaller than the bin sizes in use and much smaller than the size
of the expected signal on the magnitudes due to dust (see
\citealp{mollenhoff06}).

\section{Analysis}
\label{analysis}
The MGC structural parameters we require are the {\sc gim2d} total
luminosity, the bulge-to-total flux ratio ($B/T$), the redshift, the
redshift quality, the disc inclination and the bulge and disc
colours. Of these parameters only the disc inclination requires some
adjustment as these were derived under the assumption that galaxy
discs are infinitely thin (i.e., $\cos(i)=b/a$, where $a$ and $b$ are
the major and minor axes, respectively). In reality discs exhibit a
finite thickness preventing $b$ from reaching a value of zero. To
accommodate disc thickness we adjust our inclinations according to
Hubble (1926): $\cos^2(i) = [\cos^2(i_{\mbox{\sc
gim2d}})-Q^2]/(1-Q^2)$, where $Q$ is the ratio between the disc
scale-height and the major axis, and is set here to $0.074$ (Xilouris
et al.\ 1999).

In the following we will often use the term `inclination' in
connection with bulges. The `bulge inclination' is simply the
inclination of the associated disc.

To derive the internal attenuation--inclination relation [i.e.,
$\Delta M^*$ vs.\ $1-\cos(i)$] we iteratively follow the procedure
outlined below for both bulges and discs. Iteration is required
because any $B/T$ and magnitude cuts should be based on the intrinsic
values rather than the apparent values:
\begin{enumerate}

\item Extract all discs and bulges
  and apply the current best estimates of the disc and bulge
  attenuation--inclination corrections (no correction is applied for
  the first iteration).

\item Recompute all $B/T$ luminosity ratios.

\item Select components with $\bmgc < 20$~mag and whose parent
  galaxies have $B/T < 0.8$.\footnote{The fixed cut at $B/T < 0.8$ is
  necessary as the post-{\sc gim2d} processing of our catalogue (see
  Allen et al.\ 2006, fig.~14) replaces all systems with higher $B/T$
  values with single S\'ersic-only fits (as is common practice in
  detailed surface photometry), thereby re-defining them as
  ellipticals.}

\item Derive the luminosity distribution (LD) using step-wise maximum
  likelihood (SWML) for all discs and bulges with low inclination
  [$1-\cos(i) < 0.3$].

\item Fit Schechter functions to the LDs (using only data down to $M_B
  < -17$~mag) in order to obtain the global value of the faint-end
  slope, $\alpha$, for discs and bulges of low inclination.

\item Derive the disc and bulge LDs for the full sample in uniform
  $\cos(i)$ intervals.

\item Fit Schechter functions ($M_B < -17$~mag) with $\alpha$ fixed to
  the global value from step (v).

\item Plot the recovered $M^*$ (turn-over luminosity) versus
  $1-\cos(i)$.

\item Determine the new inclination corrections by fitting equation
  (\ref{dm_i}) below to the bulge and disc data using the
  Levenberg-Marquardt $\chi^2$ minimisation method.\footnote{During
  the fitting process the highest inclination bin is ignored as the
  accuracy of the bulge-disc decomposition is susceptible to
  break-down at this limit and, as we shall see later, this bin
  remains incomplete.}

\item Repeat the above until convergence.

\end{enumerate}

In deriving the LDs using SWML we follow Driver et al.\ (2007) (for
full details of the SWML method consult Efstathiou, Ellis \& Peterson
1988). Briefly, we assume that discs evolve according to $L_z \propto
(1+z)^{-1}$ and we use the globally derived k-corrections for the disc
components throughout.  For the bulges we adopt a milder evolution of
$L_z \propto (1+z)^{-0.5}$ and use a fixed red bulge k-correction of
$k(z) = 3.86z+12.13z^2-50.14z^3$ which is the best fit to an Sa
$15.0$~Gyr spectrum (see Poggianti 1998). For further details and
justification of these choices see Driver et al. (2007).  

Crucial to a correct implementation of SWML is the specification of
the appropriate flux limit to which each galaxy could have been
observed. The flux limit of the MGC spectroscopic survey was
$\bmgc=20$~mag, in Galactic-extinction-corrected Kron magnitudes, and
normally this would be the appropriate limit. However, we have since
revised our photometry by replacing Kron with profile-extrapolated
{\sc gim2d} magnitudes. The difference between these is generally
small (see fig.~4 of Allen et al.\ 2006) but it necessitates the
introduction of an individual magnitude limit for each galaxy. More
significant is that in implementing our dust attenuation correction
these magnitude limits must be further adjusted (in effect the
correction is analogous to a revision of the photometry). Hence the
appropriate magnitude limit to which each galaxy could have been
observed is given by: $B_{\rm lim} = 20 + M_B^{\rm T}(${\sc gim2d}$) -
M_B^{\rm T}({\rm Kron}) + M_B^{\rm C}({\rm dust\ corrected}) -
M_B^{\rm C}$, where the superscripts T and C refer to `total' and
`component', respectively, and $20$~mag is the sample's original limit
in total-galaxy, dust-uncorrected Kron magnitudes. Due to the
bulge-disc decomposition it is possible, in fact frequently the case,
that a galaxy component may lie below its parent galaxy's flux
limit. In these cases the component is rejected. Keeping these
components, i.e.\ applying the flux limits only to the galaxies, but
not their components, would introduce a bias: for example, a bulge
with some apparent magnitude below its parent galaxy's limit would
still be included in the sample if it were from a low-$B/T$ system but
not if it were from a high-$B/T$ system. Applying the flux limits to
the components significantly reduces the sample size but ensures that
it remains unbiased. A by-product is that all remaining components
will be of high signal-to-noise.

For step (ix) above we arbitrarily choose a power-law to parameterise
the attenuation--inclination relation:
\begin{equation}
\label{dm_i}
M^*_i-M^*_0 = k_1 [1-\cos(i)]^{k_2},
\end{equation}
where $M^*_i$ refers to the turn-over magnitude at inclination $i$ and
$M^*_0$ refers to the face-on turn-over magnitude.

\begin{figure*}
\centering\includegraphics[width=\textwidth]{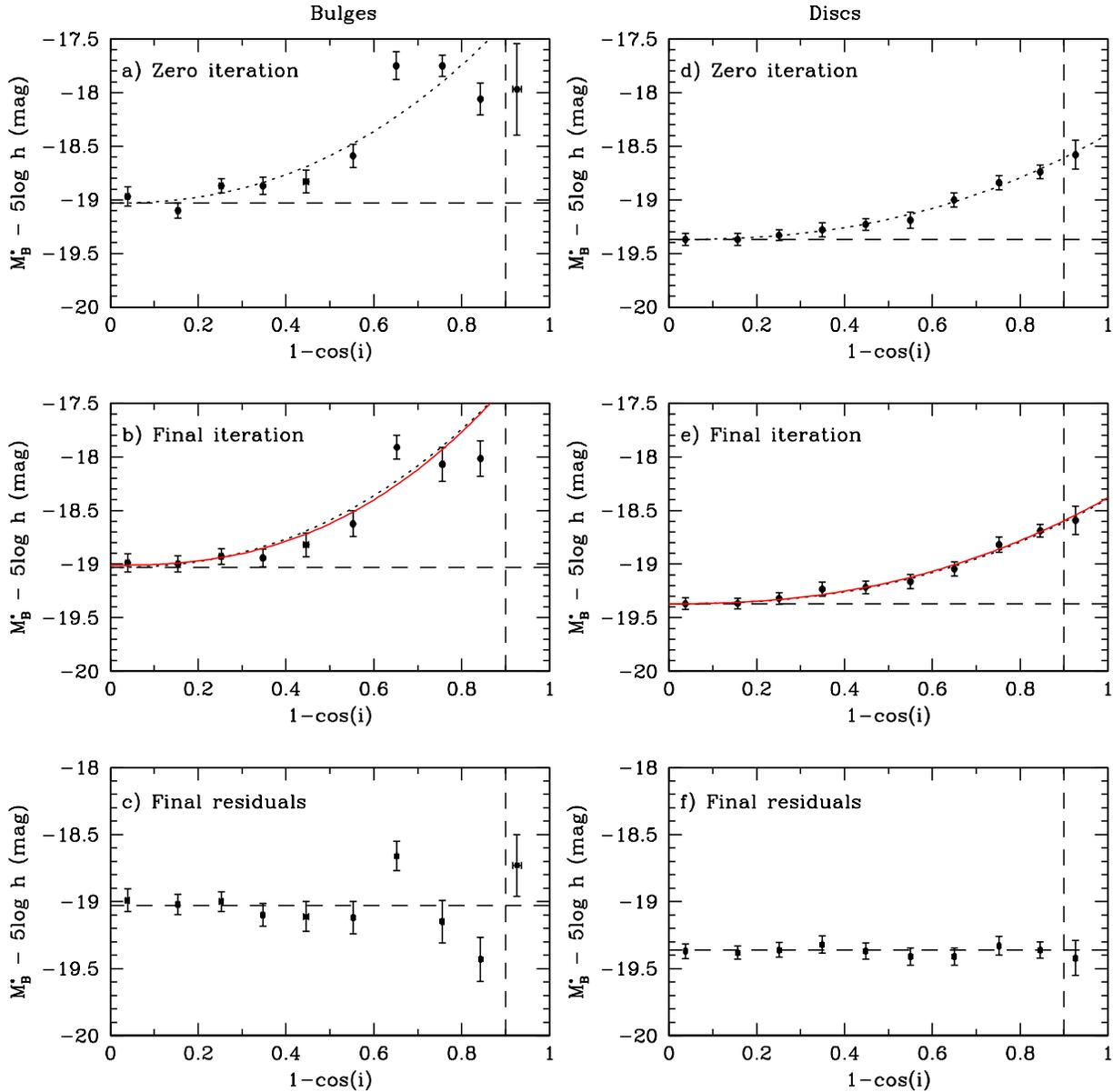}
\caption{Panels (a) and (d) show the derived $M^*$ values for bulges
and discs as a function of $1-\cos(i)$ based on the raw data. All
components are drawn from galaxies with $B/T < 0.8$ and have
luminosities $M_B < -17$~mag. The dotted line shows the best fit of
equation (\ref{dm_i}) to these data, using only the points at $1-\cos(i)
< 0.9$ (marked by the vertical dashed line). Panels (b) and (e) show
the final results after applying the inclination corrections and
re-deriving the attenuation--inclination relations repeatedly until
stable solutions are reached. The solid lines show the final fits to
these data, while the dashed lines are the same as in panels (a) and
(d). The disc relation is barely changed by the iteration and the
bulge relation has evolved only mildly. The lower panels (c and f)
show the residuals between the data points and the solid line fits.
\label{iters}}
\end{figure*}

\begin{figure}
\centering\includegraphics[width=8.0cm]{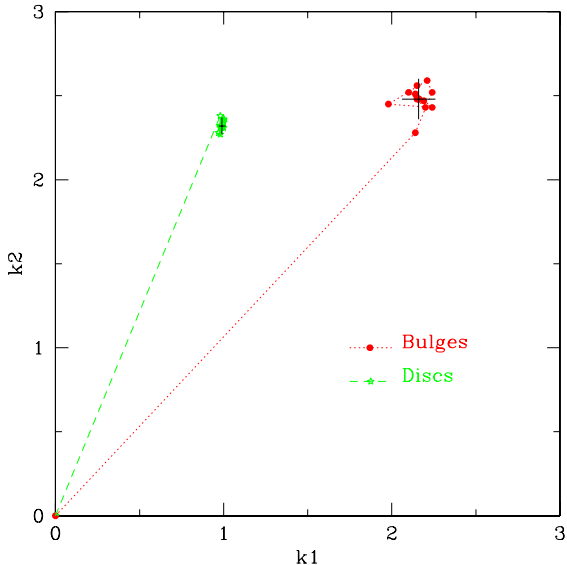}
\caption{The paths of convergence for the iterative fitting of
  equation (\ref{dm_i}) to the disc and bulge data as indicated. It is
  clear that the disc solution is extremely stable with the bulge
  solution drawing out a larger error distribution in the two fitted
  parameters. The solid crosses mark the final solutions and
  associated errors.
\label{soln}}
\end{figure}

Fig.~\ref{iters} shows the initial and final iterations for bulges
(panels a and b) and discs (panels d and e), where the components have
been drawn from galaxies with $B/T < 0.8$ and have $M_B <
-17$~mag. From Fig.~\ref{iters} we can see that both bulge and disc
magnitudes are severely underestimated in edge-on systems and a
significant correction is required.
The final results (after $14$ iterations) for the
attenuation--inclination relations are:
\begin{equation}
\label{dm_i_d}
(M^*_i-M^*_0)_{\rm disc} = (0.99\pm0.02)[1-\cos(i)]^{(2.32\pm0.05)}
\end{equation}
and
\begin{equation}
\label{dm_i_b}
(M^*_i-M^*_0)_{\rm bulge} = (2.16\pm0.1)[1-\cos(i)]^{(2.48\pm0.12)}.
\end{equation}
These relations are shown as solid lines in the central panels of
Fig.~\ref{iters}. We also show the initial relations (before any
iteration) as dotted lines. We can see that the iteration process
actually has a fairly small effect on both the disc and bulge
solutions. Fig.~\ref{soln} shows the convergence path for the disc and
bulge solutions. The discs converge almost instantly whereas the bulge
solution shows more variation. This in part reflects the larger disc
sample but may also indicate more noise in the bulge data as one might
expect for the less well resolved component. As the solutions are
dependent on each other it is reassuring that the disc solution is
robust to the variations in the bulge solution. 

\subsection{Robustness checks}
As our result is fairly striking and the implications potentially
far-reaching, it is important to ensure our interpretation is correct
and not due to some systematic artifact of the bulge-disc
decomposition process. Below we outline a number of checks which we
use to reassure ourselves, and the reader, that the result is robust
and accurately described by the equations shown above.

\subsubsection{The luminosity function fits}
\begin{figure*}
\centering\includegraphics[width=\textwidth]{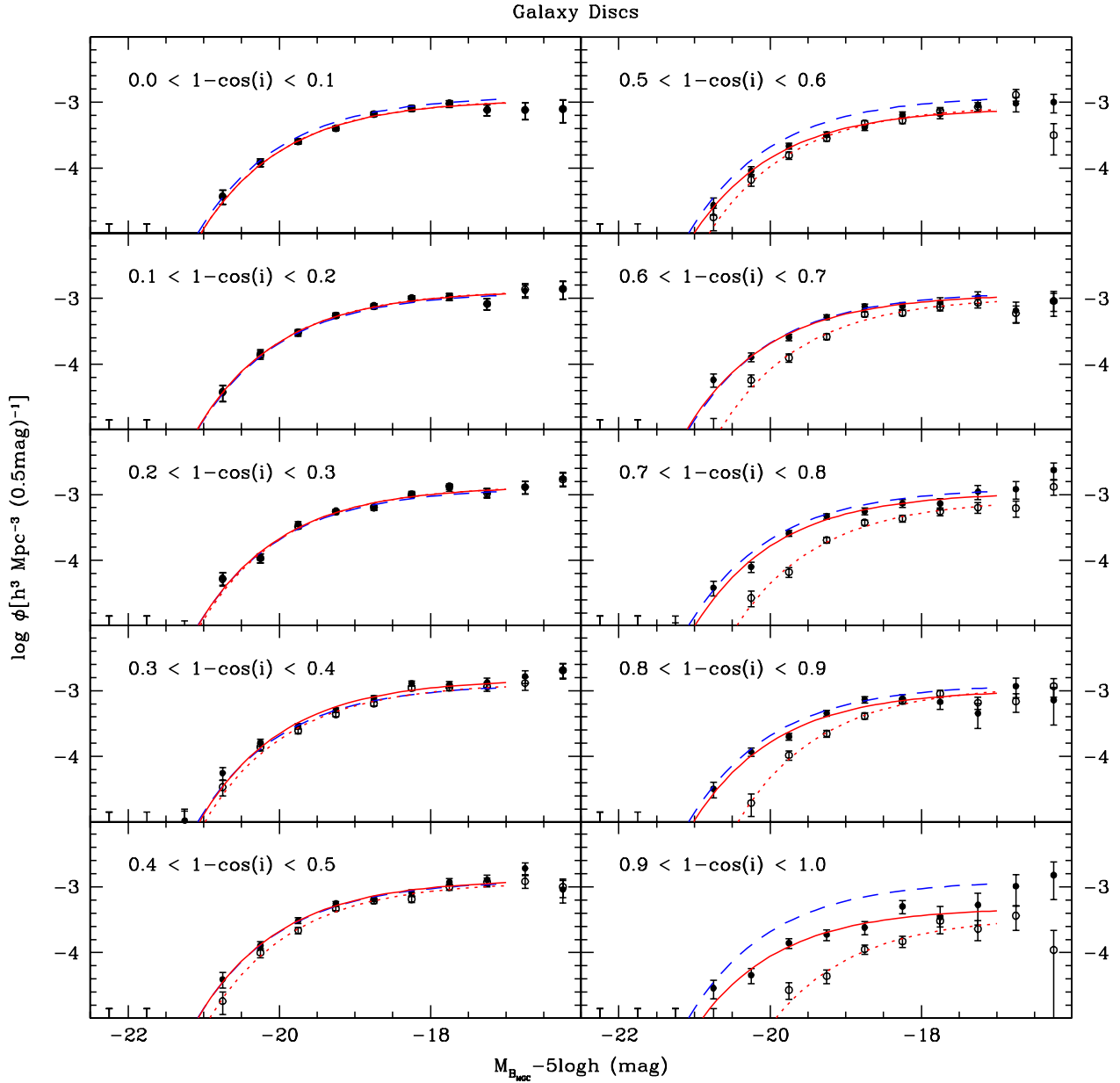}
\caption{Each panel shows the disc luminosity distribution for an
  inclination-selected sub-sample (as indicated) before (open circles)
  and after (filled circles) attenuation--inclination correction. The
  red solid and dotted lines show Schechter function fits to the
  corrected and uncorrected data, respectively. For reference, we
  also show as a blue dashed line the Schechter function derived from
  all discs with $1 - cos(i) < 0.3$, scaled down by a factor of $3$.
\label{disclf}}
\end{figure*}

\begin{figure*}
\centering\includegraphics[width=\textwidth]{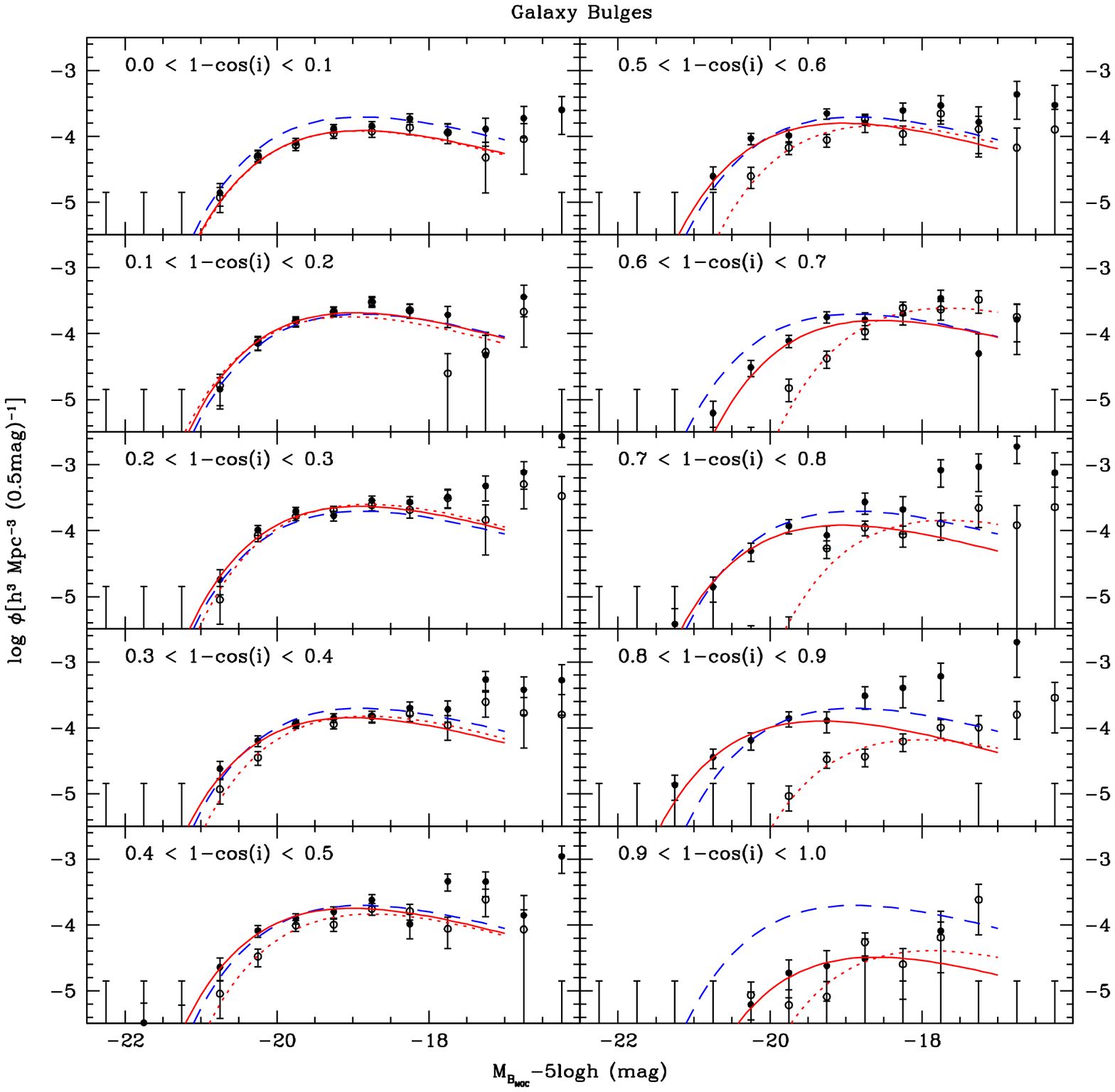}
\caption{Each panel shows the bulge luminosity distribution for an
  inclination-selected sub-sample (as indicated) before (open circles)
  and after (filled circles) attenuation--inclination correction. The
  red solid and dotted lines show Schechter function fits to the
  corrected and uncorrected data, respectively. For reference, we
  also show as a blue dashed line the Schechter function derived from
  all bulges with $1 - cos(i) < 0.3$, scaled down by a factor of $3$.
\label{bulgelf}}
\end{figure*}

Figs.~\ref{disclf} and \ref{bulgelf} show the individual luminosity
functions for discs and bulges for each $1-\cos(i)$ interval before
(dotted lines) and after (solid lines) implementing the dust-induced
inclination correction. To provide a reference we show as long dashed
lines the luminosity functions derived from bulge or disc components
with $1-\cos(i) < 0.3$, scaled down by a factor of $3$ to account for
the $3$ times larger range in $1-\cos(i)$. In general the luminosity
distributions are well behaved and the Schechter function fits are all
good. The uncorrected data (open circles) show a significant and
progressive shift to fainter $M^*$ values at higher inclinations. The
corrected data (filled circles) show that the $M^*$ values are now
consistent, as required by the fitting process. However, we can also
see that the normalisations are also consistent (cf.\ dashed
comparison line) both for the bulge and disc populations. The only
significant inconsistency is in the final, highest inclination bin
where the normalisations are low, suggesting some residual
incompleteness. As the highest inclination bin was not used in the
fitting process this does not affect our attenuation--inclination
solutions.

\subsubsection{The $1-\cos(i)$ distribution}
\label{cosisec}
\begin{figure*}
\centering\includegraphics[width=\textwidth]{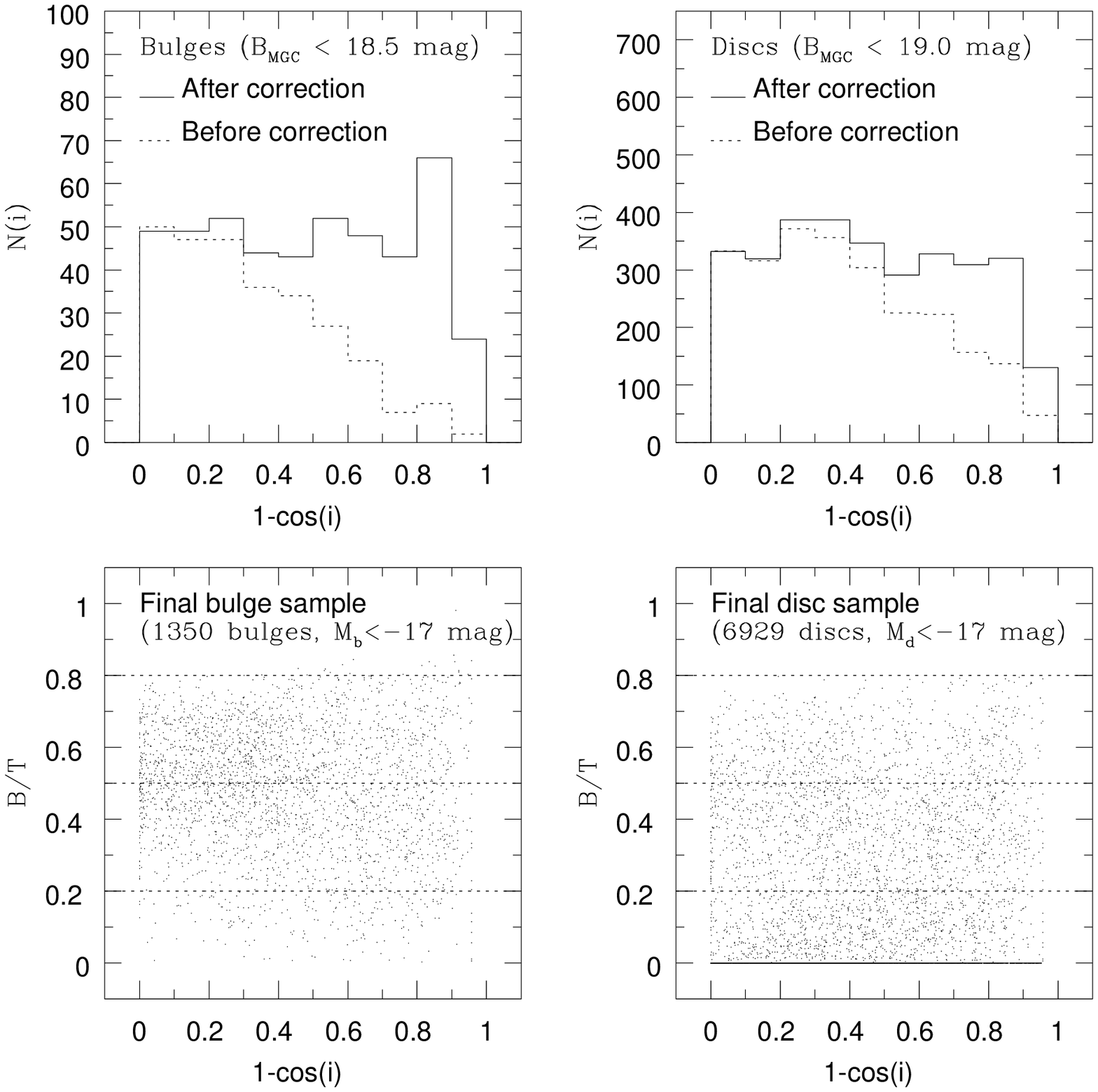}
\caption{The upper panels show the $1-\cos(i)$ distributions of bulges
(left) and discs (right) before (dotted lines) and after (solid lines)
applying the inclination dependent magnitude correction. In the lower
panels we plot $B/T$ versus $1-\cos(i)$ for the corrected data.
\label{cosi}}
\end{figure*}

The upper panels of Fig.~\ref{cosi} show the $1-\cos(i)$ distributions
before and after implementing the attenuation--inclination
corrections. For a randomly orientated sample of thin discs the
$1-\cos(i)$ distributions should be constant. Note that the magnitude
limits used in these plots are dictated by the nominal limit of the
MGC survey ($\bmgc = 20$~mag) minus the maximum bulge and disc
corrections (i.e., $\sim 1.5$ and $1.0$~mag, giving limits of $18.5$
and $19.0$~mag respectively). Fainter than these limits our samples
become incomplete, with the incompleteness depending on inclination.

From Fig.~\ref{cosi} we can see that the raw data (dotted lines)
show a strongly skewed distribution implying significant
incompleteness towards the high-inclination end, providing
corroborating evidence of the severe effect of dust attenuation.
Furthermore, bulges show a more extreme skew than discs implying that
a larger correction is required for the bulges as shown by equations
(\ref{dm_i_d}) and (\ref{dm_i_b}). After implementing the corrections
described above, both distributions are now flat. We consider both the
flatness of the $1-\cos(i)$ distribution and the uniformity of the
normalisations of the Schechter function fits from Figs.~\ref{disclf}
and \ref{bulgelf} to be extremely strong independent evidence that our
derived corrections are caused by dust attenuation and that equations
(\ref{dm_i_d}) and (\ref{dm_i_b}) are correctly accounting for this
effect. Note that the $1 - \cos(i)$ distributions also show some
residual incompleteness in the highest inclination bin, potentially
requiring the final space density of bulges and discs to be increased
by factors of $1.05$ and $1.06$, respectively.

\subsubsection{Solutions for restricted $B/T$ ranges}
\label{rcbt}
\begin{figure*}
\centering\includegraphics[width=\textwidth]{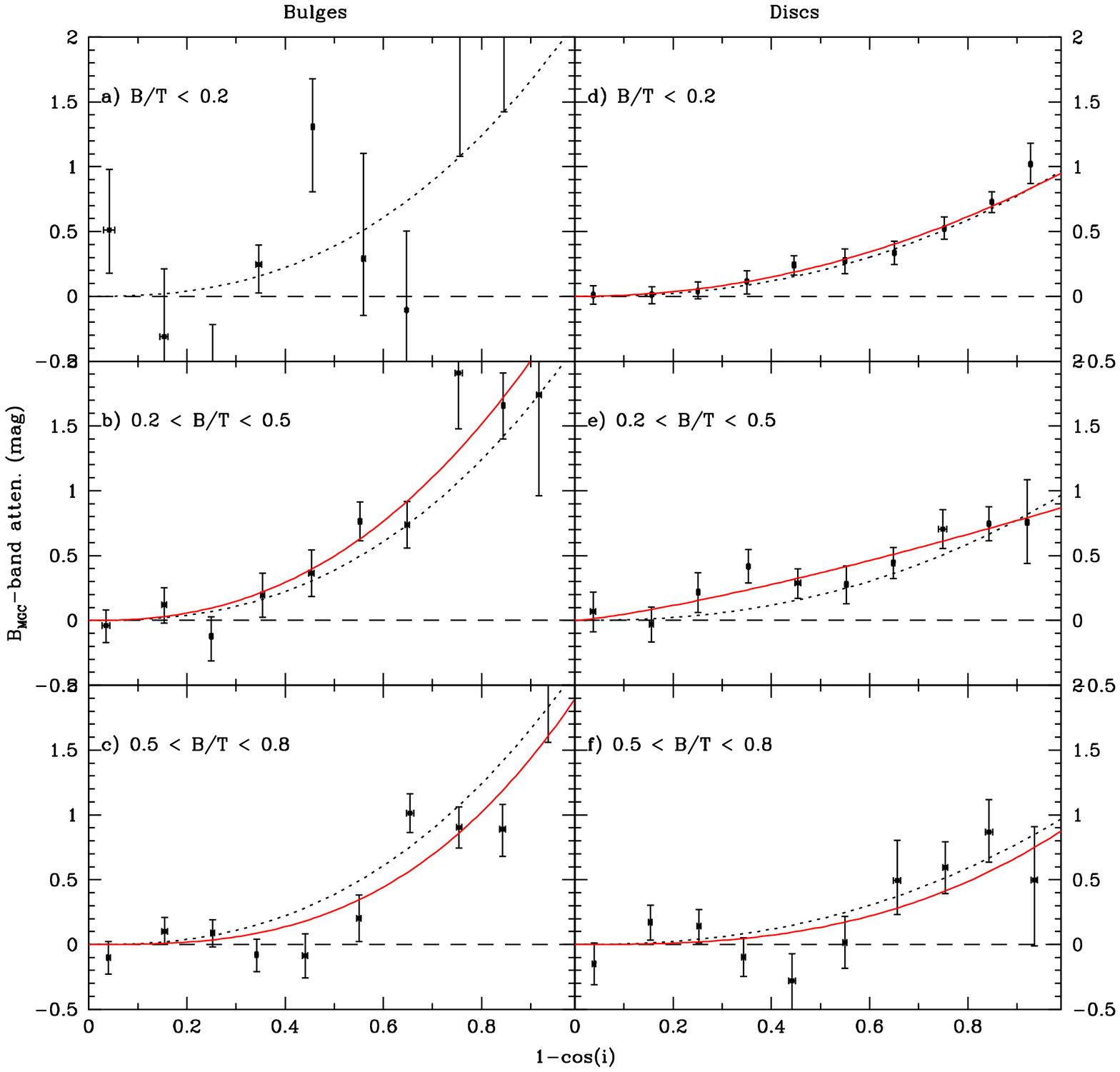}
\caption{The left and right panels show the bulge and disc
attenuation--inclination corrections derived from various
$B/T$-selected sub-samples as indicated. The dotted lines show the
solutions derived for the full samples and the solid lines show the
final fits for the specified $B/T$ interval. 
\label{bt}}
\end{figure*}

An obvious concern is that the empirical correction may depend on the
bulge-to-total flux ratio, $B/T$. Fig.~\ref{bt} shows the trends for
three different $B/T$ ranges as defined by the horizontal bands shown
in the lower panels of Fig.~\ref{cosi}. To construct these plots we
initially adopted the solutions derived from the full sample and then
followed the iterative procedure outlined in Section~\ref{analysis}.
The data for low $B/T$ bulges (Fig.~\ref{bt}a) is
particularly sparse (see also lower left panel of Fig.~\ref{cosi}) and
therefore not particularly informative, while the worst statistics for
discs are found in the high $B/T$ sample (panel f). Hence we will only
compare the results from the intermediate and high $B/T$ bulge
sub-samples (panels b and c) and from the low and intermediate $B/T$
disc sub-samples (panels d and e).

The disc correction seems reasonably insensitive to $B/T$. For bulges
there may be a tendency towards a shallower attenuation--inclination
relation for higher $B/T$ systems.  Three potential explanations for
such a trend immediately come to mind: (i) an intrinsic variation in
the opacity of discs with $B/T$; (ii) a variation in the size of
bulges relative to the scale-length of the dust disc (such that the
effective radius of the bulge becomes comparable to the scale-length
of the dust in high $B/T$ systems); and (iii) a systematic error in our
bulge-disc decompositions such that flux is transfered from bulges to
discs preferentially for lower $B/T$ systems.

The uniformity of the disc relation argues against (i). One might also
expect it to rule out (iii). However, although the transfer of flux
from bulges to discs in low $B/T$ systems can significantly modify the
flux of the bulge, it will actually have a much smaller impact on the
disc fluxes (because the systems have low $B/T$). In
Section~\ref{model} we will show that the dust model prediction for
the low $B/T$ disc sample actually reproduces the bulge trend for the
high $B/T$ sample reasonably well. This suggests that (ii) is unlikely
to be the correct explanation. We therefore conclude that we {\em may}
be slightly underestimating the flux of the bulges of low $B/T$
systems due to the bulge-disc decomposition, preferentially
re-assigning flux from the bulge to the disc at higher inclinations,
although we cannot rule out a real phenomenon, not predicted by
models.

Arguably, the above implies that it may be better to construct bulge
luminosity functions from face-on samples 
as opposed to implementing the inclination correction. The disc sample
on the other hand is robust with respect to variations in $B/T$,
indicating that dust properties scale with the disc, as one would
expect, and are entirely independent of the bulge.\footnote{Note that
we have already applied a colour cut to our bulge sample to remove
contaminating blue pseudo-bulges.} For the discs our correction is
therefore applicable over the full $B/T$ range. Hence we will continue
to use the bulge correction based on the full sample.

\subsubsection{Comparison of attenuation-corrected LFs and uncorrected 
face-on LFs} 
\begin{figure}
\vspace{-3.0cm}
\centering\includegraphics[width=\columnwidth]{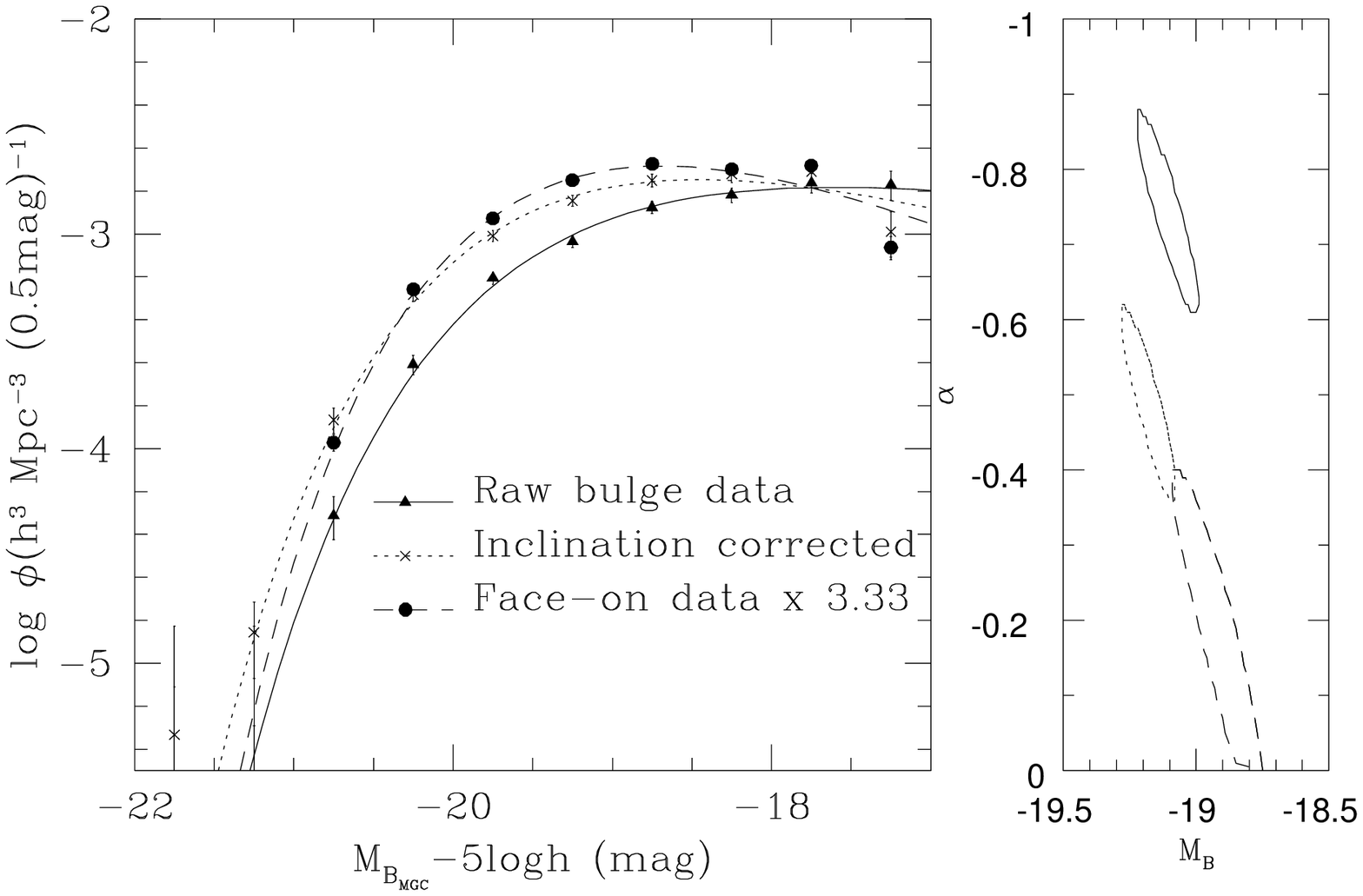}
\vspace{-3.0cm}\\
\centering\includegraphics[width=\columnwidth]{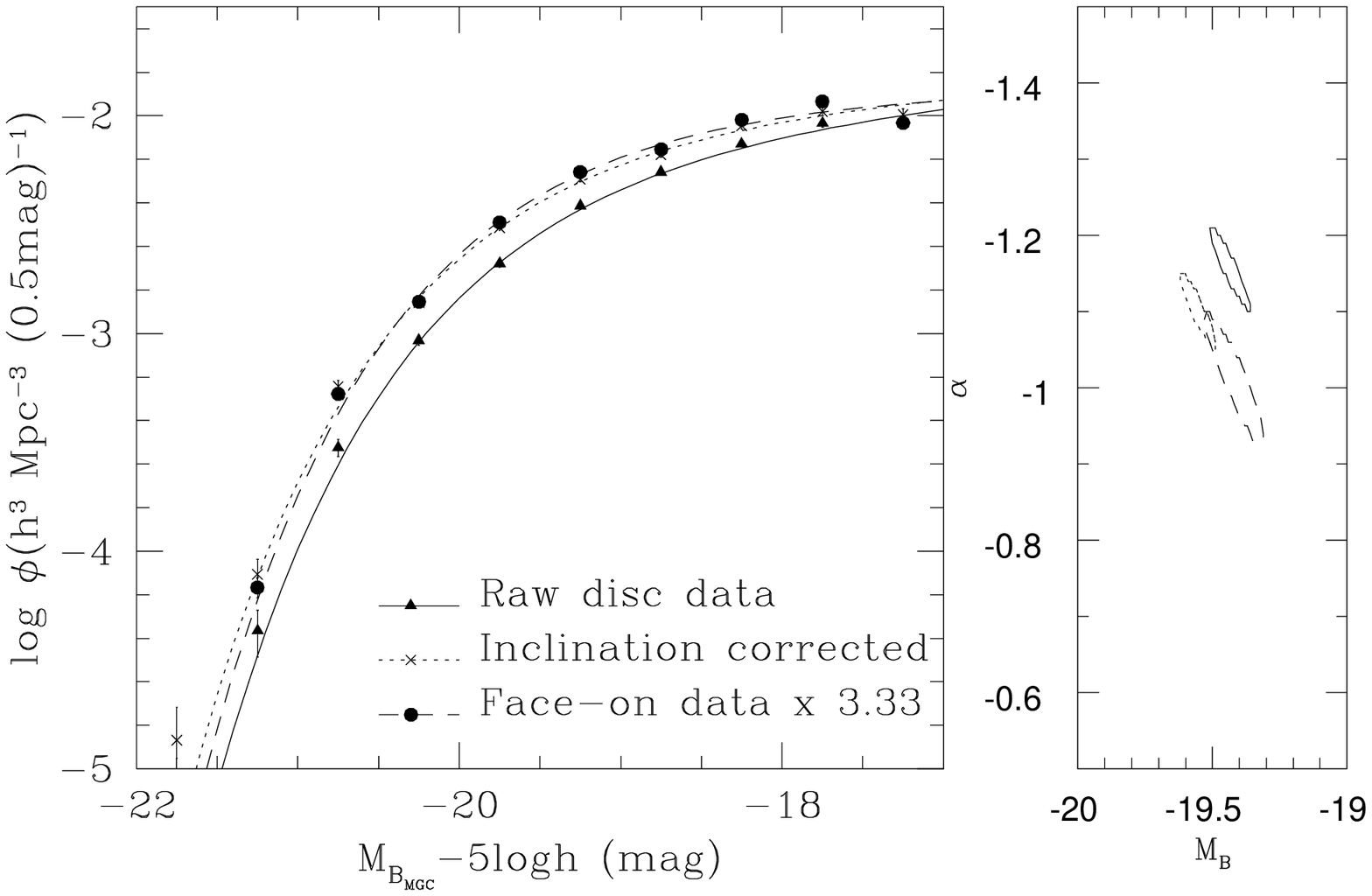}
\caption{The upper panel shows the bulge luminosity distribution and
  function before (solid line) and after (dotted line) applying the
  attenuation--inclination correction. The dashed line shows the
  luminosity function derived only from face-on [$1-\cos(i) < 0.3$]
  data and scaled up by a factor of $3.33$. The lower panel shows the
  same for discs. The right panels show the $1\sigma$ error contours
  in the $M^*$--$\alpha$ plane.
\label{lfcomp}}
\end{figure}

A further check is to re-derive the bulge and disc luminosity
distributions without applying the inclination dependent dust
correction but using face-on systems only. These should agree with the
full-sample, corrected distributions provided the normalisations are
adjusted appropriately (although the errors of the scaled-up face-on
sample are expected to be higher). Fig.~\ref{lfcomp} shows the results
which are also tabulated in Table~\ref{tbl-1}. Rows 4 and 5 show the
results for the disc luminosity functions derived from data restricted
to $1-\cos(i) < 0.3$ and scaled up by a factor of $3.33$. These
face-on estimates essentially circumvent the entire empirical
attenuation--inclination fitting process. The two luminosity function
estimates agree well with just a slight offset in
normalisation.\footnote{The inclination corrected luminosity functions
have been scaled slightly to account for the residual incompleteness
factors derived in Section~\ref{cosisec}.} Based on the $1\sigma$
error contours (right panels in Fig.~\ref{lfcomp}) the face-on data
agree with the inclination corrected data and both are formally
inconsistent with the results from the full, uncorrected sample. Note
that the error contour for the face-on sample is larger than the
others as the sample is of course smaller.

\subsubsection{GIM2D Robustness}
\begin{figure}
\centering\includegraphics[width=\columnwidth]{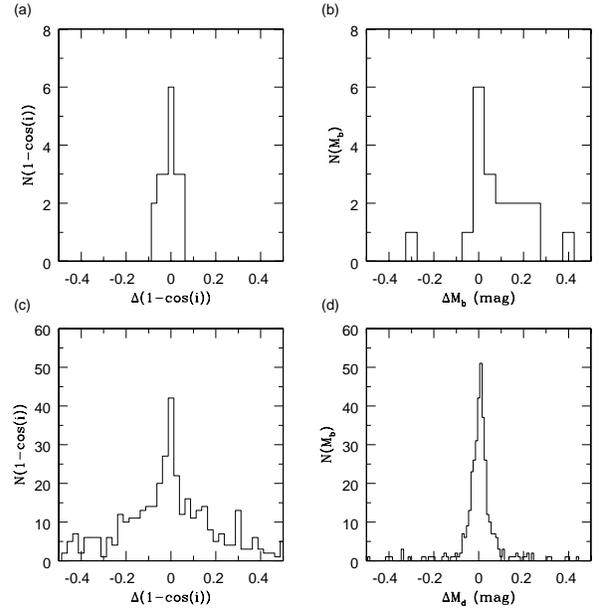}
\caption{The stability of the key {\sc gim2d} measurements of
$1-cos(i)$ (left panels) and component magnitude (right panels) for
bulges (upper panels) and discs (lower panels) for which repeat
measurements with {\sc gim2d} exist. Note that the adopted inclination
for bulge components is taken from their associated discs.
\label{repeats}}
\end{figure}

The general robustness of {\sc gim2d} has been extensively verified
via a number of studies, in particular Simard et al.\ (2002), who use
detailed simulations to verify the accuracy of the code. Comparisons
have also been made against other similar codes (e.g., Pignatelli,
Fasano \& Cassata 2006). Generally {\sc gim2d} has no systematic bias,
except perhaps with the recovery of large S\'ersic indices ($n > 4$).
Overall it is generally considered a robust and reliable code. Here we
do not repeat these studies however we do address the credibility of
our measurements via repeat analyses for a subset of galaxies which
lie in the overlap regions (see Liske et al.\ 2003). From the overlap
regions we have 682 repeat observations for which identical bulge disk
decompositions analyses and logical filtering (see Allen et al.\ 2006)
have been carried out. In this paper we are only using high
signal-to-noise components where the component magnitude is brighter
than $20.0$ mag, the absolute component magnitude is brighter than
$-17.0$ mag, and where the apparent-$B/T$ ratio is less than
0.5. After implimenting these cuts in both set of analyses we are left
with a total of 20 bulges and 389 discs for which repeat {\sc gim2d}
measurements exist.  Fig.~\ref{repeats} shows the stability of the key
measurements ($1-\cos(i)$ and component magnitude) --- note that the
inclination adopted for the bulge components is taken from their
associated discs.  These data confirm the broader analysis of the full
catalogue by Allen et al.\ (2006) that our {\sc gim2d} measurements
are repeatable. The component magnitudes are robust to better than
$\pm 0.1$ mag (for both bulges and discs) and that the $1-\cos(i)$
measurement is robust to $\sim 0.05$ for systems with high
signal-to-noise bulges and $\sim 0.1$ for disc systems.  As the error
in $1-\cos(i)$ is comparable to our bin size, for discs, this may
partially explain the smooth trend where scattering between bins will
correlate both the measurements and errors. It is worth noting that
the accuracy of the disc inclination appears more robust for those
systems with a high signal-to-noise bulge.  This is to be expected as
the $B/T$ selection ensures that only very high signal to noise discs
contribute to the bulge repeats naturally leading to a more accurate
inclination measurement. Our conclusion is that our {\sc gim2d}
results are robust as demonstrated by the repeatability of the
measurements. However we cannot rule out certain limitations in the
{\sc gim2d} software, for example non-exponential discs and disc
truncation issues which are not as yet incorporated in the {\sc gim2d}
package.

\subsubsection{Looking at images}
\begin{figure*}
\vspace{-2.0cm}
\centering\includegraphics[width=\textwidth]{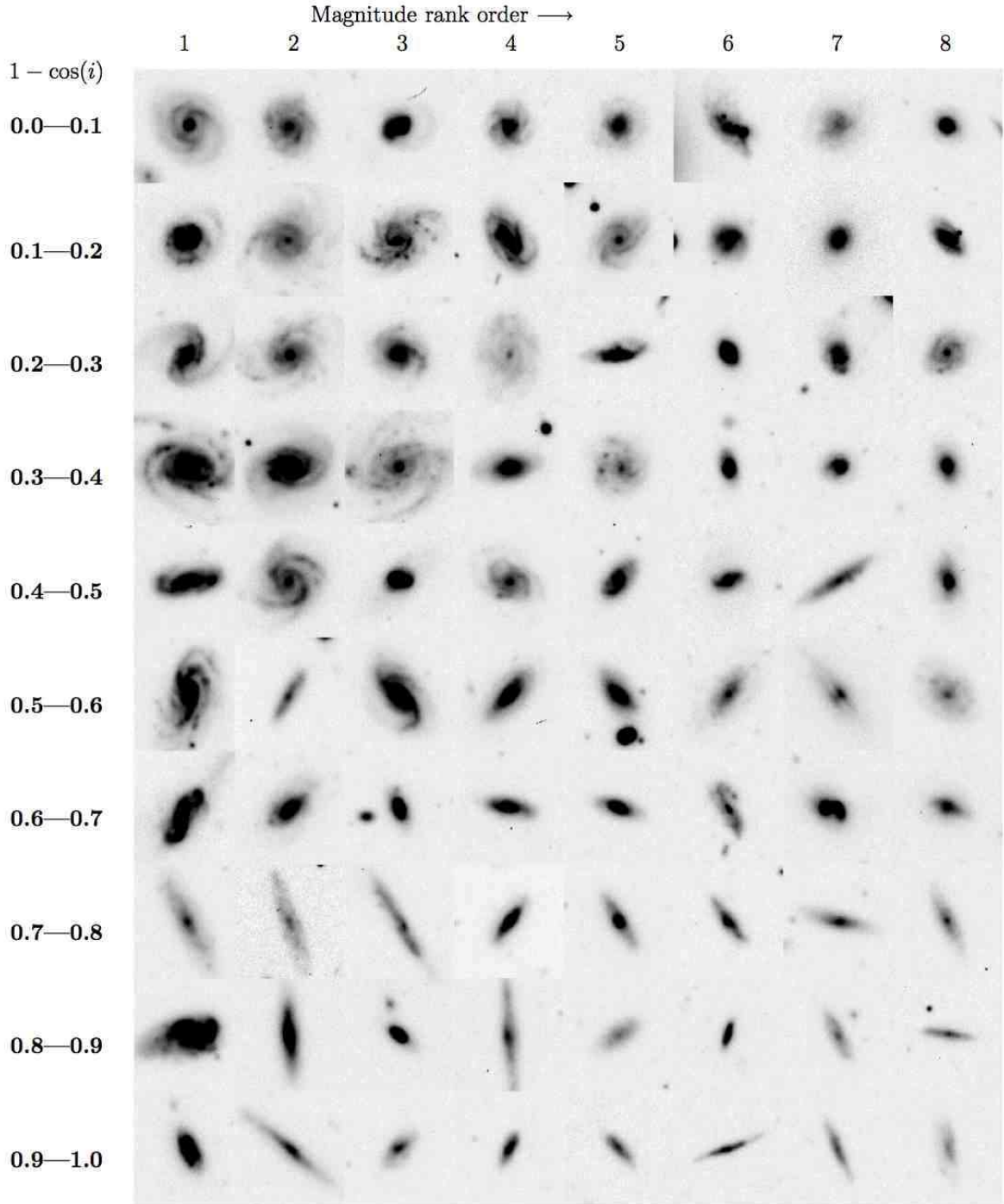}
\vspace{-2.0cm}
\caption{For each $1-\cos(i)$ bin along the vertical axis we show MGC
  postage stamp images of the eight brightest galaxies in a narrow
  redshift interval ($0.05 < z < 0.07$) and with $B/T < 0.2$. Within
  each bin the galaxies are arranged along the horizontal axis
  according to their magnitude rank. On average, and in the absence of
  dust, the galaxies should have the same intrinsic luminosity around
  the $L^*$ value, irrespective of their inclination. However, one can
  see a general diminishing of flux towards higher inclination.
\label{plate}}
\end{figure*}

A result as strong as that revealed above should be detectable by
directly inspecting the imaging data. Fig.~\ref{plate} shows galaxies
with $B/T < 0.2$ drawn from a narrow redshift interval ($0.05 < z <
0.07$). In each $1-\cos(i)$ bin (vertical axis) we show the eight
brightest galaxies, arranged along the horizontal axis by decreasing
luminosity. All galaxies are displayed at the same contrast levels,
and even though trends will be to some extent masked by cosmic
variance it is apparent that there is progressively less flux as one
looks down the sequence in inclination. Clearly some objects lie in
erroneous positions. For example there is a highly inclined galaxy in
the $1-\cos(i) = 0.4$ to $0.5$ bin and highly asymmetric systems are
also seen in inappropriate inclination bins. We estimate, roughly,
that we have a $10$ per cent failure rate in our {\sc gim2d}-derived
structural catalogue but no systematic bias sufficiently large to
dominate our result.

\section{Modelling the face-on attenuation}
\label{model}
In the previous section we obtained a purely empirical
attenuation--inclination relation. However, this relation does not
provide the face-on attenuation, nor does it give direct information
about the dust mass which is needed to account for the observed rise
in attenuation with inclination. However, both these quantities can be
estimated using a model incorporating a given geometry for the
distributions of dust and luminosity, since such models predict both
the face-on attenuation and the rise in attenuation with inclination
as a function of disc opacity. To derive these we adopt the model of
Popescu et al.\ (2000), which uses geometries for stars and dust which
can reproduce the entire UV/optical/FIR/sub-mm spectral energy
distribution (SED) of nearby spiral galaxies and for which Tuffs et
al.\ (2004) tabulated the total attenuation versus inclination as a
function of central $B$-band face-on optical depth, $\tau_B^{\rm f}$,
separately for the disc and bulge components.

\begin{figure}
\centering\includegraphics[width=\columnwidth]{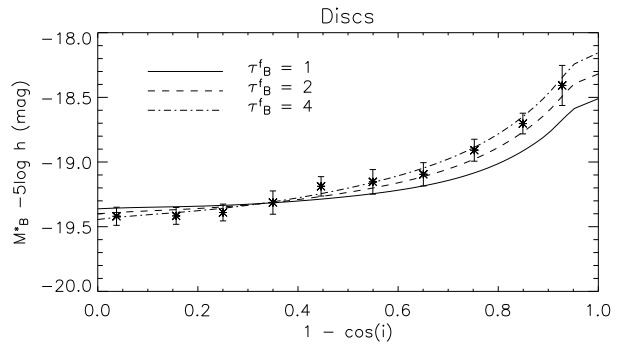}
\caption{The empirical disc attenuation--inclination relation derived
in Section~\ref{analysis} (plotted as symbols) compared to the
predictions of the dust model of Tuffs et al.\ (2004) for discs with
different central face-on $\bmgc$-band opacities, $\tau^f_B$.
\label{discmod}}
\end{figure}

\begin{figure*}
\centering\includegraphics[width=\textwidth]{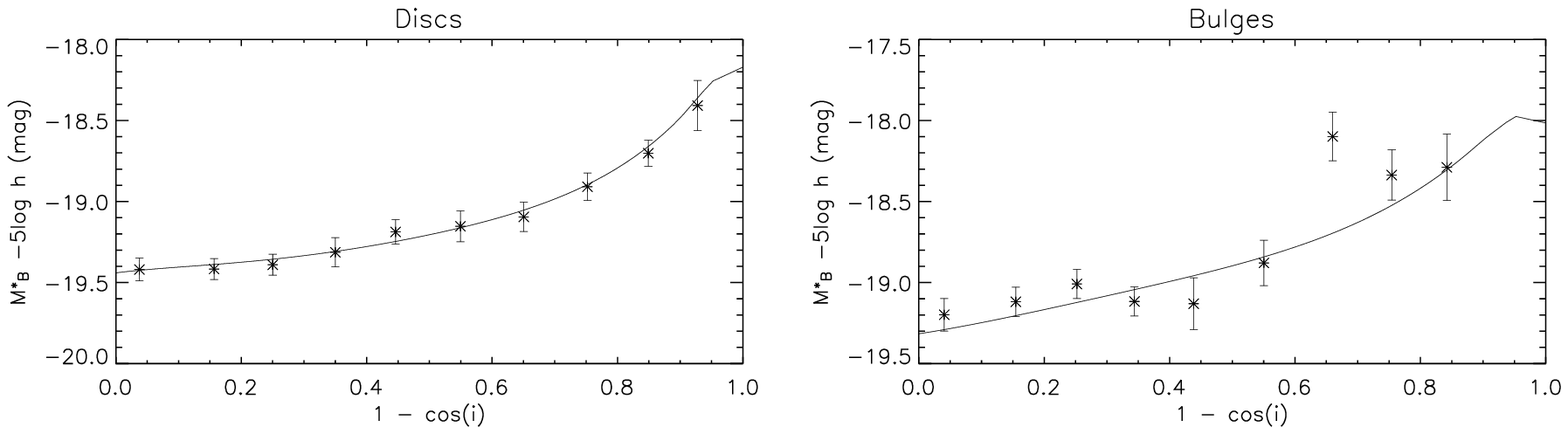}
\caption{The disc (left) and bulge (right) empirical
attenuation--inclination relation derived in Section~\ref{analysis}
compared to the prediction of the dust model of Tuffs et al.\ (2004)
with a central face-on $\bmgc$-band opacity $\tau_B^{\rm f} = 3.8$, as
derived from the best fitting solution for the
attenuation--inclination relation for discs (see Fig.~\ref{discmod}).
\label{bulgemod}}
\end{figure*}

The predictions of this model for three values of central face-on
$\bmgc$-band opacities $\tau_B^{\rm f}=$1, 2 and 4 are overlaid on the
$B/T < 0.2$ disc data in Fig.~\ref{discmod}. We only consider disc
dominated galaxies here in order to minimise any contamination from
bulges arising from any biases in the bulge-disc decomposition due to
the effect of dust itself. One can see that the model predicts the
data rather well for $\tau_B^{\rm f}= 4$. The best fit is $\tau_B^{\rm
f}=3.8 \pm 0.7$. From this value we can now derive the $\bmgc$ face-on
attenuation for discs of $(0.20 \pm 0.04)$~mag and for bulges of
$(0.84 \pm 0.10)$~mag.

This level of attenuation implies that the central regions of disc
systems are optically thick, as also found by Shao et al.\ (2007)
based on an analysis of spiral galaxies (without bulge-disc
decomposition) drawn from the Sloan Digital Sky Survey (SDSS). To put
this in context, consider the case of a geometrically infinitely thin,
optically thick disc. In this case one expects all of the light from
the far side of the bulge to be entirely blocked. This would result in
a reduction of the bulge's flux by a factor of two, i.e.\ by
$0.75$~mag. We find a slightly higher value of $0.84$~mag which
reflects the physical thickness of the dust layer. For an infinitely
thin disc, one also expects the bulge flux to increase again sharply
in the edge-on case where one can see both halves of the bulge. This
does not happen in our models, again because of the thickness of the
dust layer. Instead, the attenuation decreases only slightly as the
edge-on case is approached. The high level of face-on
attenuation of bulges derived from our models is quite insensitive to
the assumed distribution of unattenuated stellar light in the centre,
for example it does not change much when changing the S\'ersic index of
the bulge. The biggest uncertainty which could potentially affect our
conclusion, that more than half of the light from the bulge is blocked
by the dust in the disk, concerns whether or not this dust has a hole
in the centre. We do not believe this to be the case since imaging
observations of dust emission from nearby face-on spiral galaxies (at
linear resolutions of a few hundred parsecs or better) show no
evidence for an incomplete dust disk in the central regions occupied
by the bulge (Haas et al.\ 1998 and Gordon et al.\ 2006 for M31; Hippelein et
al.\ 2003 and Hinz et al.\ 2004 for M33, Popescu et al.\ 2005 for M101).

Fig.~\ref{bulgemod} shows the empirically determined
attenuation--inclination curve for the bulges of bulge-dominated
galaxies, together with the model prediction for $\tau^f_B=3.8$. The
model is broadly consistent with the data. As predicted by Tuffs et
al.\ (2004), the curve is steeper for bulges than for discs, even at
low inclinations. It is worth highlighting that the models used here
incorporate only one free parameter, $\tau_B^{\rm f}$, fitted to the
discs which provides a satisfactory fit to the full trend for both
discs and bulges simultaneously. This suggests that the large scale
geometry of dust in the MGC galaxies relative to the stars
is well described by that adopted {\em a priori} by the model. It also
follows that the basic model assumption, namely that the effective
radius of the bulge is much smaller than the scale-length of the dust
disc, is indeed valid for the bulk of galaxies in the sample,
irrespective of the value of $B/T$ (cf.\ the discussion in Section
\ref{rcbt}). It also begs the question as to whether entirely embedded
bulges may exist in some systems.

We also note that our conclusion that the measured
attenuation--inclination curves require galaxies with fairly opaque
central regions is qualitatively consistent with the lack of a strong
increase in the central surface brightness of the galaxies depicted in
Fig.~\ref{plate} as one progresses from face-on to edge-on
systems. Such an increase is predicted for optically thin galaxies
\citep{mollenhoff06}.
 

\section{The cosmic stellar luminosity, mass and dust densities}
\label{cosmic}
\subsection{Intrinsic luminosity functions and the cosmic stellar luminosity
  density}
\label{ilf}
Having derived the empirical attenuation--inclination relation and the
model dependent face-on attenuation for both discs and bulges we are
now in a position to derive the luminosity functions of the
pre-attenuated $\bmgc$ flux produced by the total stellar population
(i.e., before any attenuation occurs). To do this we follow Driver et
al.\ (2007) after first applying our attenuation--inclination relation
to all discs and bulges derived from systems with post-corrected $B/T$
values ranging up to the relatively large value of $0.8$. The validity
of also correcting bulge dominated systems is demonstrated by our own
data, which show that galaxies with $0.5 < B/T < 0.8$ are attenuated
almost as much by dust as galaxies with $0.2 < B/T < 0.5$ (see
Figs.~\ref{bt}b and c), and by the
local examples of the Centaurus A and Sombrero galaxies. Furthermore,
many S0 galaxies obey the FIR--radio correlation, showing that they
have similar dust opacities in relation to their star-formation
activity as spiral galaxies \citep{bally89}. Note that it is not
clear, at this stage, whether the S0s in our sample are predominantly
found among our high-$B/T$ or elliptical populations. Most likely they
fall in both. Higher spatial resolution imaging is required before
they can be identified with any confidence and hence at this point we
cannot make any reliable statements about their dust content.

Fig.~\ref{lfs} and Table~\ref{tbl-1} show the final results based on
both the inclination and the combined inclination and face-on
corrections. In making these corrections we assumed that opacity does
not depend on luminosity, i.e.\ we applied the same correction for all
galaxies. The recent study by Shao et al.\ (2007) suggests this
assumption is reasonable as they find minimal change in opacity with
luminosity.

We note that all three Schechter parameters have been substantially
modified by taking into account the effects of dust. This is
especially true for bulges, where $L^*$ is increased by a factor of
$2.3$, $\phi^*$ is increased by $58$ per cent and $\alpha$ is increased
by $0.19$. This change in $\alpha$ is caused by objects previously
below the flux cut-off being dust corrected into the sample. The
luminosity densities corresponding to the various Schechter function
fits are also given in Table~\ref{tbl-1}.

\subsection{The cosmic stellar mass densities}
\label{csmd}
\begin{figure}
\centering\includegraphics[width=\columnwidth]{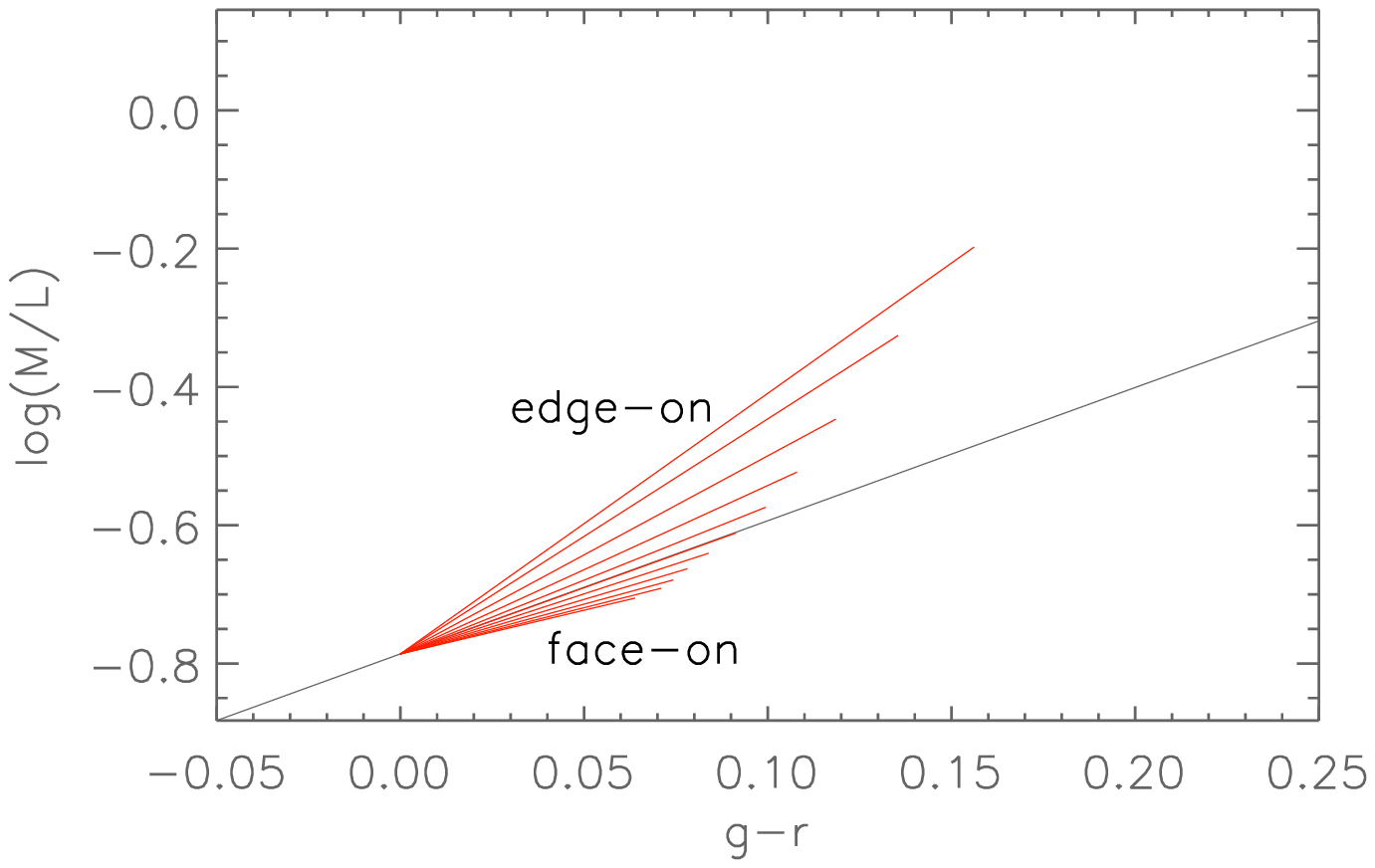}
\centering\includegraphics[width=\columnwidth]{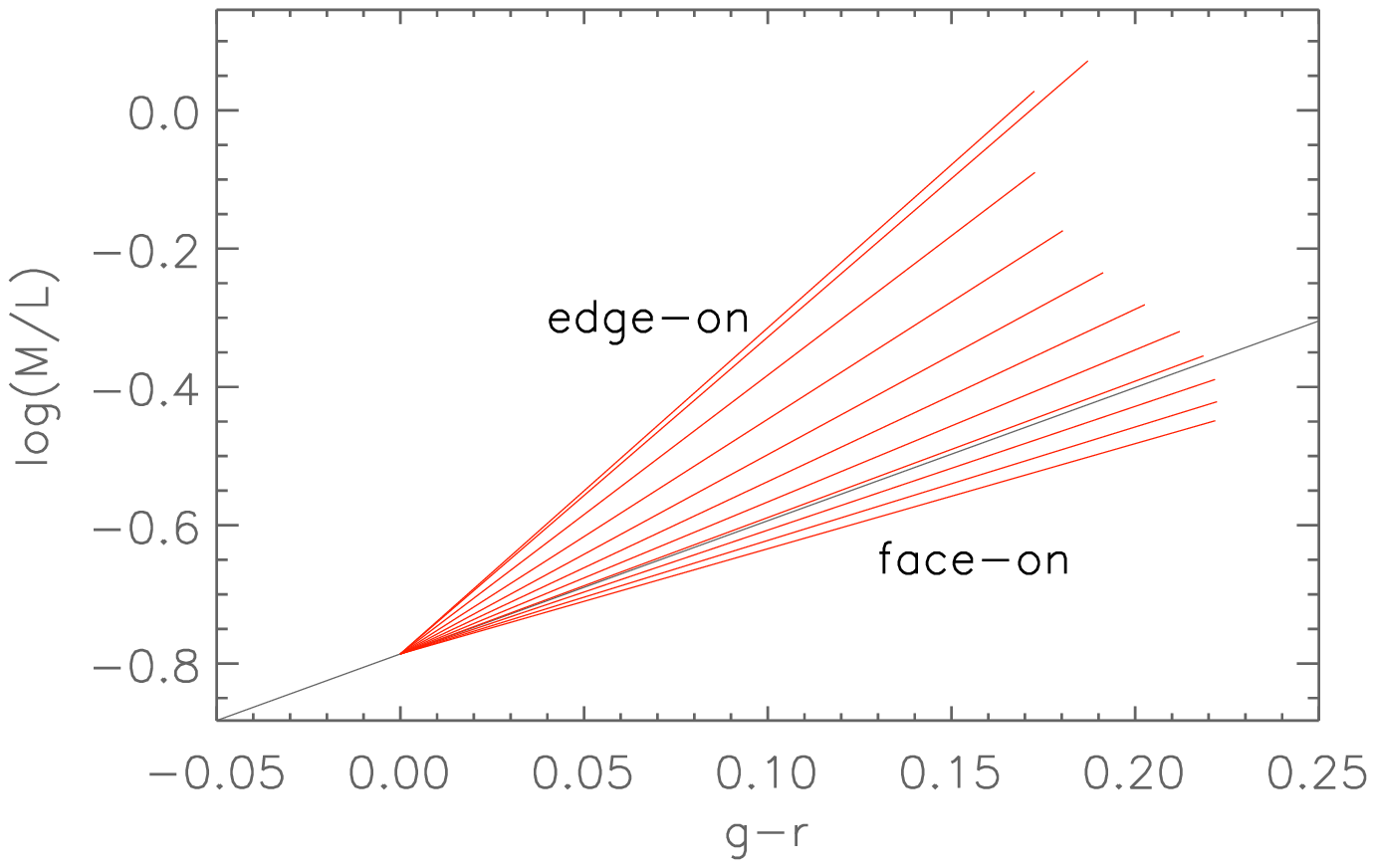}
\caption{This figure shows the predicted relation (black line) between
intrinsic mass-to-$B$-band light ratio and intrinsic $g-r$ colour,
derived by Bell \& de Jong (2001) for a fixed total stellar mass. We
over-plot $11$ vectors in red showing the shift of a point on the
correlation due to dust for inclinations $1-\cos(i)=0, 0.1, \dots,
1$. The upper and lower panels show the disc and bulge cases,
respectively. The vectors are calculated from the model of Tuffs et
al.\ (2004) for $\tau^f_B=3.8$.
Vectors which move away from the fixed mass locus imply a significant
change in the galaxy's total stellar mass, which appears to be the
case for edge on systems. \label{bdj}}
\end{figure}

\begin{figure}
\centering\includegraphics[width=\columnwidth]{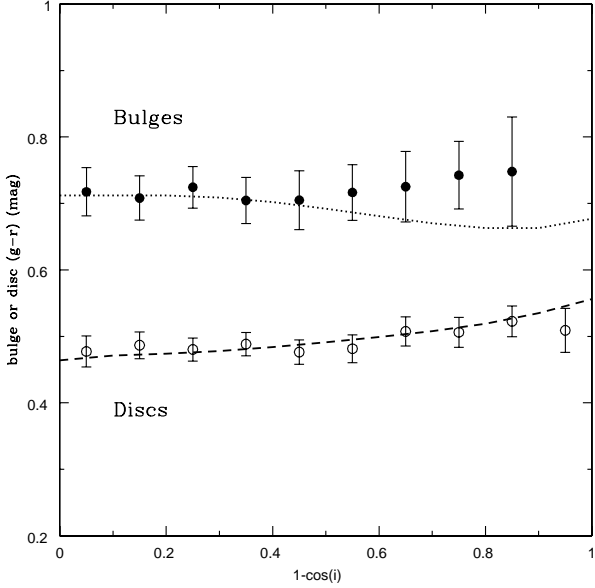}
\caption{The component-$(g-r)$ colour verses inclination for bulges
and discs derived empirically (data points) and predicted by our dust
models (lines) . \label{grinc}}
\end{figure}

In addition to the luminosity densities we also derive the stellar
mass densities (Table~\ref{tbl-1}, column 6). We follow the procedure
outlined in Driver et al.\ (2006) which uses the $(g-r)$ disc or bulge
colour to derive a mass-to-light ratio based on the prescription given
by Bell \& de Jong (2001). The adopted equation (see Driver et al.\
2006) is:
\begin{equation}
\label{stm}
\mathcal{M} = 10^{[1.93(g-r)-0.79]} 10^{-0.4(M_B-M_{\odot})},
\end{equation}
where $\mathcal{M}$ denotes stellar mass, and $g-r$ and $M_B$ refer to
attenuation free quantities. Since we have already corrected the
magnitudes for attenuation we must now also correct the colour for the
effect of dust before we can apply this formula. Fig.~\ref{bdj} shows
the combined correction for attenuation and reddening for an example
point on the mass-to-$B$-band light ratio versus $g-r$ colour
relation, separately for discs and bulges. This correction was derived
from Tables 4 and 6 of Tuffs et al.\ (2004) for $\tau^f_B=3.8$ and for
$11$ inclinations corresponding to $1-\cos(i)=0, 0.1, \dots, 1$. One
sees a strong dependence of the corrections on inclination. The colour
corrections are listed in Table~2 and are derived from our disc
opacity constraint on the general dust models presented in Tuffs et
al.\ (2004). Empirically we cannot determine the absolute colour
corrections to corroborate these values, however we can compare the
{\it inclination}-dependent component. Fig.~\ref{grinc} shows the mean
component $(g-r)$-colour from within each inclination bin. As can be
seen from a comparison of the models and data the {\it
inclination}-dependent colour correction is actually quite weak but in
full agreement for discs across the entire inclination range and for
bulges with $1-\cos(i)<0.6$.

Bell \& de Jong argue that the stellar masses derived from their
relation are robust against the effect of dust provided the dust
vectors are parallel to the relation. The reason is that the
under-prediction of stellar mass arising from the attenuation in
luminosity will be compensated by the over-prediction of stellar mass
arising from the reddening in colour. Inspection of Fig.~\ref{bdj}
shows that whereas this balance applies for face-on systems (both for
discs and bulges) this is not the case for highly inclined systems.
Both bulges and discs at higher inclinations show a systematically
higher ratio of attenuation to reddening. This situation arises
because a larger fraction of the lines of sight through both bulges
and the central regions of discs at all inclinations are optically
thick, and in the optically thick limit attenuation becomes saturated
at a high level, exhibiting only a small variation with wavelength
(see Tuffs et al.\ 2004). The overall consequence is that stellar
masses will be under-predicted for a randomly oriented distribution by
use of the Bell \& de Jong relation if dust is not taken into account.

The corrections shown in Fig.~\ref{bdj} were applied to each object
individually and the stellar masses were extracted using equation
(\ref{stm}). These estimates for the mass densities (as well as for the
stellar luminosity) will be valid if the efficiency of absorption of
$\bmgc$-band photons by $M^*$ galaxies is representative of the galaxy
population at large.

\begin{figure}
\centering\includegraphics[width=\columnwidth]{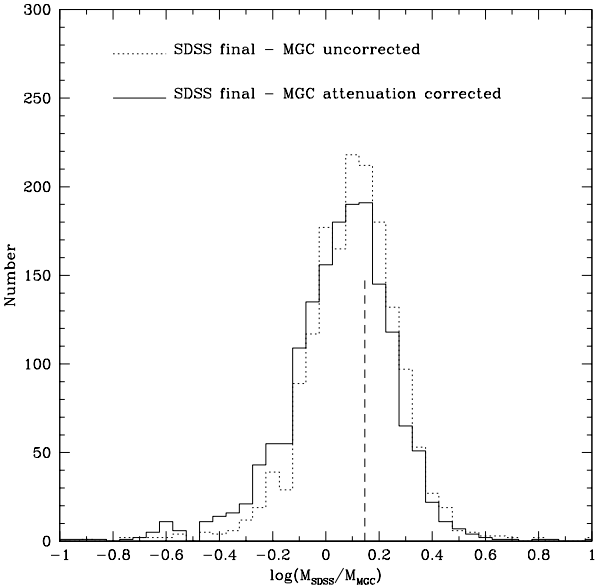}
\caption{Comparison of the final dust corrected SDSS stellar masses
(Kauffmann et al.\ 2003) and the MGC total-galaxy stellar masses for
MGC data with (dotted line) and without (solid line) our dust
correction. The corrected MGC data is shifted to higher masses,
relative to SDSS values and the final median offset is a factor of
$\sim$$1.18$ (i.e., $0.073$~dex). \label{smass}}
\end{figure}

In order to ascertain the credibility of our stellar mass estimates we
compare our masses to those derived for SDSS galaxies by Kauffmann et
al.\ (2003). The data were kindly provided by J.~Brinchman (priv.\
comm.) and were matched to the MGC catalogue, resulting in $1855$
matched objects (using a $5$~arcsec positional tolerance). The
relatively small number of matches is due to the much brighter limit
of the SDSS spectroscopic survey compared to the MGC. The SDSS stellar
mass-to-light ratios are derived from the spectra, a dust attenuation
correction from the colours and the final masses from the $z$-band
Petrosian fluxes (incorporating the derived mass-to-light ratio and
attenuation correction).  Full details of this process are given by
Kauffmann et al.\ (2003).

Fig.~\ref{smass} shows the comparison of the matched sample both with
and without dust correction (which applies to both the SDSS and MGC
data). Note that we are comparing total-galaxy stellar masses here,
i.e.\ for two-component galaxies we have summed the MGC disc and bulge
estimates. The median stellar mass ratio is found to be $1.18$
(i.e., $0.073$~dex) with SDSS masses being systematically higher. SDSS
masses are based on a Kroupa (2001) IMF whereas the Bell \& de Jong
values are based on a Salpeter-lite IMF but these should give
comparable masses. The dispersion is reasonable, implying an
uncertainty of $\Delta \log(\mathcal{M})= \pm 0.16$ ($45$ per cent)
for individual galaxy measurements which includes a contribution from
the magnitude uncertainty. If one assumes the stellar mass error is
equally distributed between the two surveys this implies individual
mass uncertainties of: $\Delta \log(\mathcal{M}) \sim \pm 0.11$ (i.e.,
$30$ per cent). We therefore conclude that our mass estimates show a
reasonable dispersion but are still systematically lower than SDSS
estimates by $\sim 18$ per cent. Examining the data more closely one
can see a trend for more discrepant masses amongst bluer star forming
galaxies, this will be explored further in a future paper.

The stellar mass densities shown in column 6 of Table~\ref{tbl-1} are
derived from a simple sum over all bulges or discs, where each
component's stellar mass is first multiplied by a weight. This weight
is the space density at the component's luminosity (from the relevant
luminosity function) divided by the number of systems which
contributed to that luminosity interval (see Driver et al.\ 2006). The
values for all parameters (Schechter function, luminosity density and
stellar mass density) shown in Table~\ref{tbl-1} are revised from
those shown in Driver et al.\ (2007), because of the dust
correction. It is instructive to compare the final luminosity and
stellar mass densities derived from the attenuation--inclination
correction and from the rescaled face-on values. In almost all cases
we see that the rescaled values are slightly higher; this {\em may} be
a further reflection of residual incompleteness in the highest
inclination bin (cf.\ the $1-\cos(i)$ distributions of Fig.~\ref{cosi}
and the luminosity functions shown in Fig.~\ref{lfcomp}). One can
estimate a correction factor to apply to the attenuation corrected
data based on the incompleteness of extreme inclination objects. These
are $1.05$ for bulges and $1.06$ for discs resulting in final stellar
mass density estimates, at redshift zero, of:
\begin{eqnarray}
\rho({\rm discs}) & = & (4.4 \pm 0.3) \times 10^8 \, h \, M_{\odot} \, 
{\rm Mpc}^{-3} \nonumber\\
\rho({\rm bulges}) & = & (2.2 \pm 0.2) \times 10^8 \, h \, M_{\odot} \, 
{\rm Mpc}^{-3} \\
\rho({\rm ellipticals}) & = & (0.8 \pm 0.1) \times 10^8 \, h \,M_{\odot} \,
{\rm Mpc}^{-3} \nonumber
\end{eqnarray}
Hence we conclude that the combined stellar mass density based on our
empirical inclination plus model face-on correction is $\rho = (7.6
\pm 0.4 \pm 1.0) \times 10^8 \, h \, M_{\odot}$~Mpc$^{-3}$, where the
second error is due to cosmic variance.  This should be modified
downwards by about a factor of $1.1$ to get a Kroupa IMF based value
(I.~Baldry priv.\ comm.). Note that this total value also includes
the stellar mass estimate, derived in Driver et al.\ (2007), for the
low luminosity blue spheroid systems of $\rho({\rm blue\ spheroids}) =
(0.2 \pm 0.1) \times 10^8 \, h \,M_{\odot} \, {\rm Mpc}^{-3}$.

\begin{table*}
\begin{minipage}{\textwidth}
\caption{Schechter function parameters for various galaxy samples with 
varying degrees of dust attenuation corrections.
\label{tbl-1}}
\begin{tabular}{lccccc}
\hline\hline
Component & $M^* - 5 \log h$ & $\phi^*$ & $\alpha$ & $j^a$ & $\rho^b$ \\ 
& (mag) & ($10^{-2} \, h^3$ Mpc$^{-3}$ (0.5 mag)$^{-1}$) & & ($10^8 \, h\, L_{\odot}$ Mpc$^{-3}$) & ($10^8 \, h\, M_\odot$ Mpc$^{-3}$) \\ \hline
Discs:\\
uncorrected                                            & $-19.45 \pm 0.04$ & $1.8          \pm 0.1^d$  & $-1.16 \pm 0.04$ &  $1.7        \pm 0.2^d$  & $3.8        \pm 0.4^d$ \\
inclination corrected                                  & $-19.56 \pm 0.04$ & $2.1(2.2)^c   \pm 0.1^d$  & $-1.10 \pm 0.03$ &  $2.1(2.2)^c \pm 0.2^d$  & $4.7(5.0)^c \pm 0.5^d$ \\
inclination \& face-on corrected                       & $-19.76 \pm 0.04$ & $2.1(2.2)^c   \pm 0.1^d$  & $-1.11 \pm 0.03$ &  $2.6(2.7)^c \pm 0.2^d$  & $4.1(4.4)^c \pm 0.3^d$ \\ 
$3.33 \times (1-\cos(i) < 0.3)$                        & $-19.44 \pm 0.07$ & $2.7          \pm 0.2^d$  & $-1.02 \pm 0.06$ &  $2.3        \pm 0.5^d$  & $5.3        \pm 1.2^d$ \\ 
$3.33 \times (1-\cos(i) < 0.3)$ \& face-on corrected   & $-19.64 \pm 0.07$ & $2.6          \pm 0.2^d$  & $-1.04 \pm 0.05$ &  $2.7        \pm 0.5^d$  & $4.7        \pm 0.9^d$ \\ \hline
Bulges:\\
uncorrected                                            & $-19.11 \pm 0.07$ & $0.65         \pm 0.05^d$ & $-0.75 \pm 0.08$ &  $0.37         \pm 0.06^d$  & $1.6        \pm 0.3^d$ \\
inclination corrected                                  & $-19.18 \pm 0.07$ & $0.87(0.91)^c \pm 0.05^d$ & $-0.49 \pm 0.08$ &  $0.51(0.54)^c \pm 0.06^d$  & $2.3(2.4)^c \pm 0.3^d$ \\
inclination \& face-on corrected                       & $-20.00 \pm 0.07$ & $0.97(1.03)^c \pm 0.05^d$ & $-0.56 \pm 0.08$ &  $1.22(1.29)^c \pm 0.15^d$  & $2.1(2.2)^c \pm 0.2^d$ \\ 
$1-\cos(i) < 0.3$                                      & $-18.92 \pm 0.11$ & $1.23         \pm 0.07^d$  & $-0.19 \pm 0.15$ & $0.60         \pm 0.08^d$  & $2.7        \pm 0.4^d$ \\ 
$1-\cos(i) < 0.3$ \& face-on corrected                 & $-19.68 \pm 0.10$ & $1.27         \pm 0.07^d$  & $-0.19 \pm 0.15$ & $1.25         \pm 0.15^d$  & $2.1        \pm 0.3^d$ \\ \hline
Spheroids (bulges + ellipticals)$^e$:\\
uncorrected                                            & $-19.15 \pm 0.06$ & $0.99         \pm 0.05^d$ & $-0.66 \pm 0.07$ &  $0.57       \pm 0.07^d$  & $2.4        \pm 0.4^d$ \\
inclination corrected                                  & $-19.15 \pm 0.06$ & $1.23         \pm 0.05^d$ & $-0.44 \pm 0.07$ &  $0.71       \pm 0.08^d$  & $3.1(3.2)^c \pm 0.4^d$ \\
inclination \& face-on corrected                       & $-19.95 \pm 0.06$ & $1.37         \pm 0.07^d$ & $-0.76 \pm 0.05$ &  $1.7        \pm 0.2^d$  & $2.9(3.0)^c \pm 0.3^d$ \\ 
$3.33 \times (1-\cos(i) < 0.3)$                        & $-19.00 \pm 0.05$ & $1.53         \pm 0.15^d$  & $-0.31 \pm 0.07$ & $0.8        \pm 0.1^d$  & $3.5        \pm 0.5^d$ \\ 
$3.33 \times (1-\cos(i) < 0.3)$ \& face-on corrected   & $-19.75 \pm 0.04$ & $1.61         \pm 0.15^d$  & $-0.52 \pm 0.05$ & $1.6        \pm 0.2^d$  & $2.9        \pm 0.4^d$ \\ \hline
Ellipticals (no corrections)                           & $-19.02 \pm 0.11$ & $0.37         \pm 0.03^d$ & $-0.26 \pm 0.14$ &  $0.20       \pm 0.03^d$ & $0.8        \pm 0.1^d$ \\ \hline
\end{tabular}\\
$^a$The luminosity density is defined as $j=\phi^*L^*\Gamma(\alpha+2)$, 
where we use $M_\odot = 5.38$~mag. To convert from the $\bmgc$ to the
$b_{\rm J}$-band multiply all $j$ values by $1.05$.

$^b$The stellar mass density is derived using equation (\ref{stm}).
The colour corrections shown in Table~\ref{tbl-2} are only used when
calculating face-on corrected values.

$^c$All $\phi^*$, $j$ and $\rho$ values for inclination
corrected data and inclination and face-on corrected data can arguably
be scaled up by factors of $1.05$ for bulges and $1.06$ for discs if
one wishes to compensate for the apparent incompleteness in the
highest inclination bin of Fig.~\ref{cosi}.

$^d$The quoted errors on $\phi^*$, $j$ and $\rho$ are the
random errors only. Based on mock 2dFGRS NGP catalogues (Cole et al.\ 1998) we
estimate that the potential systematic error on these values due to cosmic variance amounts
to $13$ per cent.

$^e$All corrections and the $\cos(i)$ selection are only applied to the
bulges. No corrections are applied to the ellipticals.

\end{minipage}
\end{table*}

\begin{table}
\begin{center}
\caption{Inclination dependent $(g-r)$ colour corrections for bulges
and discs.}
\label{tbl-2}
\begin{tabular}{ccc}
\hline\hline
$1-\cos(i)$ & bulge $(g-r)$ corr. & disc $(g-r)$ corr. \\ \hline
0.0 & 0.222 & 0.064 \\
0.1 & 0.222 & 0.071 \\
0.2 & 0.222 & 0.074 \\
0.3 & 0.219 & 0.078 \\
0.4 & 0.212 & 0.084 \\
0.5 & 0.202 & 0.091 \\
0.6 & 0.191 & 0.099 \\
0.7 & 0.180 & 0.108 \\
0.8 & 0.173 & 0.119 \\
0.9 & 0.173 & 0.135 \\
1.0 & 0.187 & 0.156 \\ \hline
\end{tabular}
\end{center}
\end{table}

\subsection{The cosmic dust density}
The derivation of the cosmic dust density from the MGC data is less
straightforward, as there is no obvious direct physical link between
the mass of dust in a galaxy and the $B$-band luminosity. Neither is
there direct empirical information linking these two quantities over
the luminosity range covered by the MGC due to the lack of
statistically significant FIR or sub-mm measurements of galaxies less
luminous than $M^*$. Here we simply assume that there is a fixed ratio
between dust mass and $\bmgc$-band intrinsic luminosity of discs, as
adopted for the canonical galaxy model used by Tuffs et al.\ (2004)
for $M^*$, and scale the cosmic $\bmgc$-band luminosity density of
discs, as derived in Section~\ref{ilf}, by this ratio.

As it is mainly discs close to $M^*$ which dominate the total disc
luminosity density, a luminosity dependent dust mass-to-light ratio
would not actually be problematic as long as the dependency is not
extreme. In this context we note that our value for the opacity of
discs ($\tau_B^{\rm f}= 3.8\pm 0.7$) is comparable to that obtained
from detailed SED modelling of the components of stellar luminosity
absorbed by grains and re-radiated in the FIR (Popescu et al.\ 2000;
Misiriotis et al.\ 2001; Misiriotis et al.\ 2004) of well studied,
relatively luminous $L^*$ spiral galaxies. We augment this sample with
data for spiral and dwarf galaxies with {\it ISO} detections drawn
from Tuffs et al.\ (2002), for which Popescu et al.\ (2002) derive dust
masses for $29$ galaxies with FIR and $B$-band luminosity
measurements.
In these systems, the ratio of dust mass per $B$-band disc intrinsic
luminosity is found to be $(0.00196 \pm 0.00058) \, M_\odot
L_\odot^{-1}$ (note no $h$ dependence). Multiplying this ratio by the
intrinsic $\bmgc$ luminosity density of discs of $2.7 \times 10^8
\,h\, L_\odot$~Mpc$^{-3}$ then yields a value of $(5.3 \pm 1.6 \pm
0.7) \times 10^5 \, h \, M_{\odot}$~Mpc$^{-3}$ for the cosmic dust
density, where the second error is due to cosmic variance.

\subsection{Implications for the cosmic baryon budget}
Adopting a value for the total cosmic baryon density of $\Omega_{\rm
baryon} h^2 = 0.023$ (Tegmark et al.\ 2006) and $\rho_{\rm
crit}=2.7755 \times 10^{11} \, h^2 \, M_{\odot}$~Mpc$^{-3}$, we find
the fraction of cosmic baryons in dust and stars today is $(0.0083 \pm
0.0027) \, h$ per cent and $(11.9 \pm 1.7) \, h$ per cent
(Salpeter-lite), respectively. This value for the stellar baryon
fraction is marginally ($2\sigma$) higher than that previously derived
by Baldry \& Glazebrook (2003) of $(5$ -- $9) \, h$ per cent
(marginalized over a variety of IMFs). Our value is also slightly
higher than the value of $(9 \pm 1.3) \, h$ per cent derived by Cole
et al.\ (2001) from the 2MASS/2dFGRS NIR luminosity function (see also
the summary of stellar masses in Bell et al.\ 2003).  We note that
Cole et al.\ (2002) cite their values as dust free but acknowledge
that if dust attenuation is severe this could impact upon their
estimates. Using our model we can explore this by deriving dust
attenuation values for the bulge and disc components in the $K_s$-band
(see Fig.~\ref{bulgediscmodel}) which demonstrate that dust
attenuation remains non-negligible even at $K_s$.

\section{Summary and Discussion}
\label{summary}
We have demonstrated that dust attenuation is a severe issue in the
$B$-band, resulting in the magnitudes of both discs and bulge
components being severely underestimated by factors of
$0.20$--$1.1$~mag and $0.84$--$2.6$~mag respectively. The direct
implication is that only $63$ and $29$ per cent of the total $B$-band
photons produced by stars in discs and bulges, respectively, actually
make it out of the galaxy (as deduced from the luminosity densities
with and without dust corrections). The remainder are absorbed by the
dust and presumably re-radiated in the FIR. As this re-radiation is
likely to be almost perfectly isotropic and un-attenuated one might
expect the optical to FIR flux ratio to show some inclination
dependence. We find that the central face-on optical depth of discs is
$\tau^f_B =3.8\pm0.7$ which implies that discs are optically thick in
the centres. This conclusion was also reached by Shao et al.\ (2007)
who analysed the behaviour of total-galaxy magnitudes of spiral
galaxies drawn from the SDSS as a function of inclination. We find
that this conclusion holds regardless of the bulge luminosity and it
is hence independent of the bulge-to-total flux ratio. Thus, dust
appears to be inherently related to the disc, with no physical
connection to the bulges.


Our value for the opacity of discs ($\tau_B^{\rm f}= 3.8\pm 0.7$) is
comparable to that obtained from self-consistent SED modelling of the
components of stellar luminosity absorbed by grains and re-radiated in
the FIR (Popescu et al.\ 2000, Misiriotis et al.\ 2001, Misiriotis et
al.\ 2004). This value for central face-on opacity is however
significantly higher than might be expected on the basis of the
extinction measurements towards central stellar clusters in spiral
galaxies by Sarzi et al.\ (2005). As a possible explanation for this
discrepancy, we note that our measurement of $\tau_B^{\rm f}$ (like
that from the SED modelling) is derived from the integrated properties
of galaxies. It is therefore representative of the global distribution
of opacity over the whole disc, and is not sensitive to opacity along
individual lines of sight; in other words, it is not sensitive to
inhomogeneities in the dust distribution on scales of
$100$~pc. Therefore, this difference may indicate that the dust
density in the central $100$~pc of spiral galaxies may be
systematically lower than for the surrounding regions of the inner
disc, and that the central $100$~pc region may have some specific
properties, e.g.\ due to feedback in the form of dispersion of
material surrounding a central massive star cluster. Alternatively, it
may simply be that the disc population has a large range of opacities,
and that central star clusters are only clearly identifiable in discs
with relatively low opacities.

Although this work quantifies the severity of the mean dust
attenuation in galaxy discs and bulges this method cannot constrain
the galaxy-to-galaxy variation. To address this it is imperative that
pointed FIR observations of large, optically selected galaxy samples
are undertaken with sufficient depth to detect systems over the full
luminosity range sampled by surveys such as the MGC.

In terms of the total $B$-band luminosity density of the Universe,
based on luminosity function estimates, we infer that our previous
estimates (Liske et al.\ 2003, 2006; Driver et al.\ 2005, 2006, 2007)
should be revised upwards by a factor of $\sim$$1.8$. In terms of the
stellar mass density the change amounts to an upward revision of a
factor of $\sim$$1.2$. Previous estimates of the $B$-band galaxy
luminosity function which have neglected to correct for dust
attenuation, have underestimated $M^*_B$ by $\sim0.6-1$~mag. When
comparing models of galaxy formation and evolution to data one must
therefore be careful to either include realistic dust attenuation in
the models or to compare to dust corrected data as provided here.

In Driver et al.\ (2007) we reported that the stellar mass was broken
down as $60$:$27$:$13$ into discs:classical bulges:ellipticals
(ignoring the contribution from pseudo-bulges and blue ellipticals).
We can now revise these values, incorporating our dust corrections for
discs and bulges, to $59$:$30$:$11$. Hence the stellar mass in
classical bulges is a factor of $2.7$ higher than the mass in
ellipticals.

Throughout this study we have assumed ellipticals to be dust
free. Since ellipticals and bulges are known to have similar apparent
colours, the correction of our bulges for dust attenuation implies
that they will have significantly bluer intrinsic colours. It
therefore follows that they must be younger and/or metal poorer which
is consistent with the spectroscopic studies of Proctor \& Sansom
(2002) and Thomas \& Davies (2006).

We also note that the high opacities leading to our conclusion that
the bulges have intrinsically bluer colours is not inconsistent with
the observations that bulges show a small spread in apparent colours
and exhibit small colour gradients (see review by Renzini 2006). Our
dust model predicts that for the high opacities found in this paper
the variation in bulge colour due to dust is almost independent of
inclination (see also Section~\ref{csmd}), and that there should be
very little colour gradients in bulges due to dust.

Finally, our calibrated dust model predicts that dust will
significantly attenuate stellar light even at NIR wavelengths,
particularly for bulges. This is illustrated in
Fig.~\ref{bulgediscmodel} where predicted attenuation--inclination
curves are plotted for the bulge and disc components of a $\tau_B^{\rm
f}=3.8 \pm 0.7$ galaxy in the $B$, $I$ and $K_s$ bands. Hence even the
upcoming deep NIR surveys will need to manage internal dust
attenuation with care.

\begin{figure}
\centering\includegraphics[width=\columnwidth]{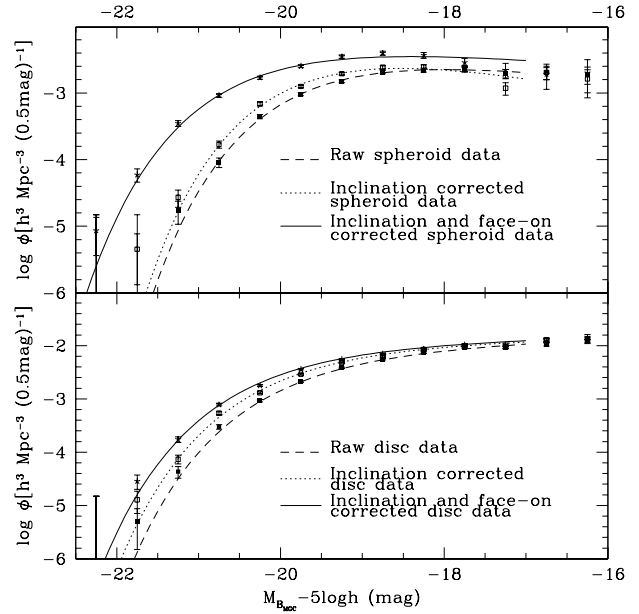}
\caption{The upper panel shows the spheroid (i.e.\ bulge+elliptical)
luminosity function before and after corrections and the lower panel
shows the disc luminosity function before and after
corrections. \label{lfs}}
\end{figure}

\begin{figure}
\centering\includegraphics[width=\columnwidth]{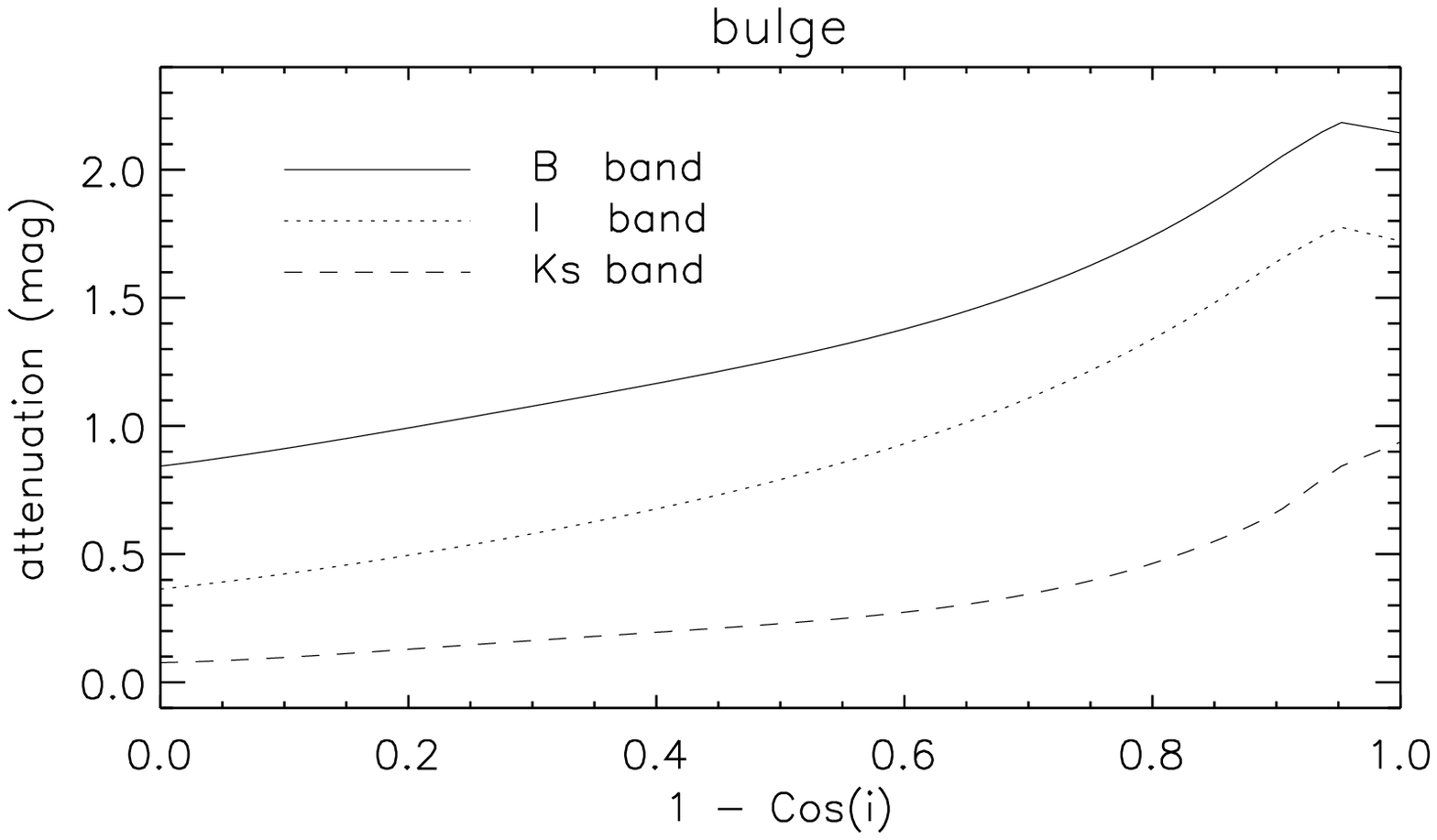}
\centering\includegraphics[width=\columnwidth]{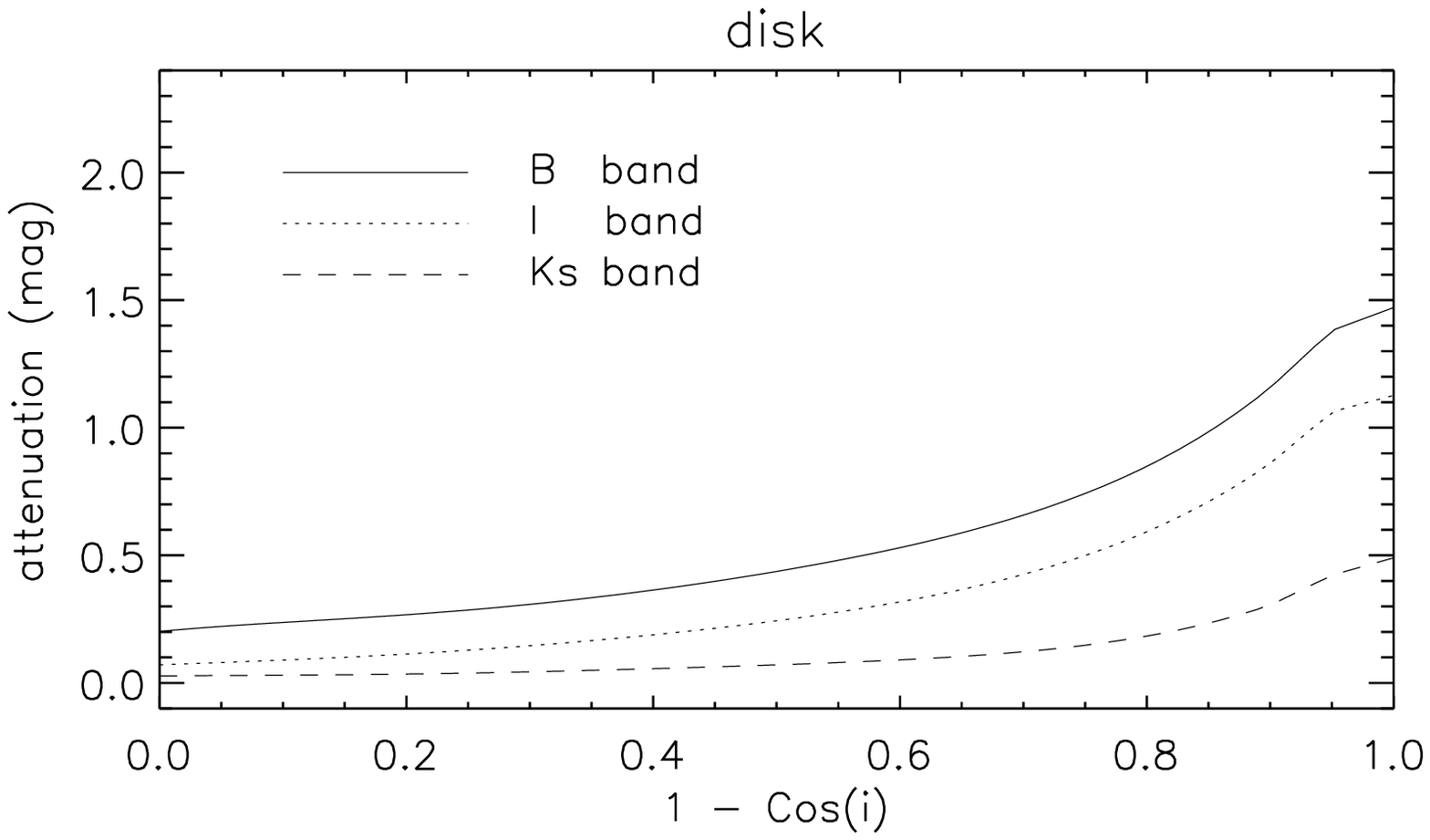}
\caption{Predicted attenuation--inclination curves in the $B$, $I$ and
$K_s$ bands for the bulge (top panel) and the disc (lower panel) of a
galaxy with a central face-on optical depth in the $\bmgc$-band of
$\tau_B^{\rm f}=3.8$.
\label{bulgediscmodel}}
\end{figure}

\section*{Acknowledgments}
We thank Jarle Brinchman for providing SDSS stellar mass data and Ivan
Baldry for comments on earlier drafts. We would also like to
acknowledge enlightening discussions with Barry Madore, Alan Dressler
and Anne Sansom. The Millennium Galaxy Catalogue consists of imaging
data from the Isaac Newton Telescope and spectroscopic data from the
Anglo Australian Telescope, the ANU 2.3m, the ESO New Technology
Telescope, the Telescopio Nazionale Galileo and the Gemini North
Telescope. The survey has been supported through grants from the
Particle Physics and Astronomy Research Council (UK) and the
Australian Research Council (AUS). The data and data products are
publicly available from http://www.eso.org/$\sim$jliske/mgc/ or on
request from J.~Liske or S.P.~Driver.

\label{lastpage}

\end{document}